\documentclass[a4paper,10pt,twocolumn]{article}

\usepackage[T1]{fontenc}
\usepackage[utf8]{inputenc}

\usepackage{authblk}
\usepackage{cite}

\usepackage{parskip}
\usepackage[top=2.0cm,bottom=2.0cm,left=2.0cm,right=2.0cm]{geometry}
\usepackage{amsmath,amssymb}

\usepackage{cool}

\usepackage{graphicx}
\usepackage{wrapfig}
\usepackage{tikz}

\usepackage[belowskip=2pt,aboveskip=15pt]{caption}

\usepackage{hyperref}
\hypersetup{
  colorlinks=false,
  pdfborder={0 0 0}
}

\title{Regularized Harmonic Surface Deformation}

\author{Yeara Kozlov}
\author{Janick Martinez Esturo}
\author{Hans-Peter Seidel}
\author{Tino Weinkauf}
\affil{Max Planck Institute for Informatics, Germany\\\vspace{1ex}\small\{ykozlov,janick,hpseidel,weinkauf\}@mpi-inf.mpg.de}
\date{}

\newcommand{\iso}{\text{iso}}
\newcommand{\conf}{\text{conf}}

\newcommand{\nvertices}{|\mathcal{V}|}

\newcommand{\ninternaledges}{|\mathcal{E}_i|}
\newcommand{\ntriangles}{|\mathcal{T}|
}

\newcommand{\transpose}{^\mathsf{T}}

\newcommand{\mathbbr}{\mathbb{R}}

\newcommand{\mySetNotation}[1]{{\mathbb{#1}}}
\newcommand{\RRSet}{{\mySetNotation{R}}}

\newcommand{\myBoldNotation}[1]{{\mathbf #1}}

\newcommand{\vc}   {{\myBoldNotation{c}}}

\newcommand{\vh}   {{\myBoldNotation{h}}}

\newcommand{\vn}   {{\myBoldNotation{n}}}

\newcommand{\vu}   {{\myBoldNotation{u}}}

\newcommand{\vx}   {{\myBoldNotation{x}}}
\newcommand{\vy}   {{\myBoldNotation{y}}}

\newcommand{\vNull}{{\myBoldNotation{0}}}

\newcommand{\mA}   {{\myBoldNotation{A}}}
\newcommand{\mB}   {{\myBoldNotation{B}}}
\newcommand{\mC}   {{\myBoldNotation{C}}}
\newcommand{\mD}   {{\myBoldNotation{D}}}
\newcommand{\mE}   {{\myBoldNotation{E}}}

\newcommand{\mG}   {{\myBoldNotation{G}}}

\newcommand{\mI}   {{\myBoldNotation{I}}}

\newcommand{\mL}   {{\myBoldNotation{L}}}
\newcommand{\mM}   {{\myBoldNotation{M}}}
\newcommand{\mN}   {{\myBoldNotation{N}}}

\newcommand{\mR}   {{\myBoldNotation{R}}}

\newcommand{\mU}   {{\myBoldNotation{U}}}

\newcommand{\mW}   {{\myBoldNotation{W}}}
\newcommand{\mX}   {{\myBoldNotation{X}}}
\newcommand{\mY}   {{\myBoldNotation{Y}}}
\newcommand{\mZ}   {{\myBoldNotation{Z}}}

\newcommand{\invn}[2][0cm]{\mathopen{}\left|\left|{#2}\parbox[h][#1]{0cm}{}\right|\right|}

\newcommand{\cE}{{\mathcal E}}

\newcommand{\cM}{{\mathcal M}}

\newcommand{\cP}{{\mathcal P}}

\newcommand{\cR}{{\mathcal R}}

\newcommand{\cT}{{\mathcal T}}

\newcommand{\cV}{{\mathcal V}}

\Style{IdentityMatrixSymb=\mI,%
       DSymb={\mathrm d},%
       IntegrateDifferentialDSymb={\mathrm d},%
       TrParen=p,%
       DetParen=p}
\Style{DDisplayFunc=outset}
\renewcommand{\Transpose}[1]{ {#1}^\Transp }

\newcommand{\T}[1]{\Transpose{#1}}

\newcommand{\Transp}{{{\mathrm T}}}

\let\oldmb\mathbold
\protected\def\mathbold{\oldmb}

\newcommand{\ie}{i.e.}

\newcommand{\eg}{e.g.}

\newcommand{\etal}{et al.\ }

\begin{document}

\DeclareGraphicsExtensions{.jpg,.pdf,.png}
\graphicspath{{fig/}}

\maketitle

\begin{figure*}
\centering%
\def\svgwidth{1\linewidth} \small 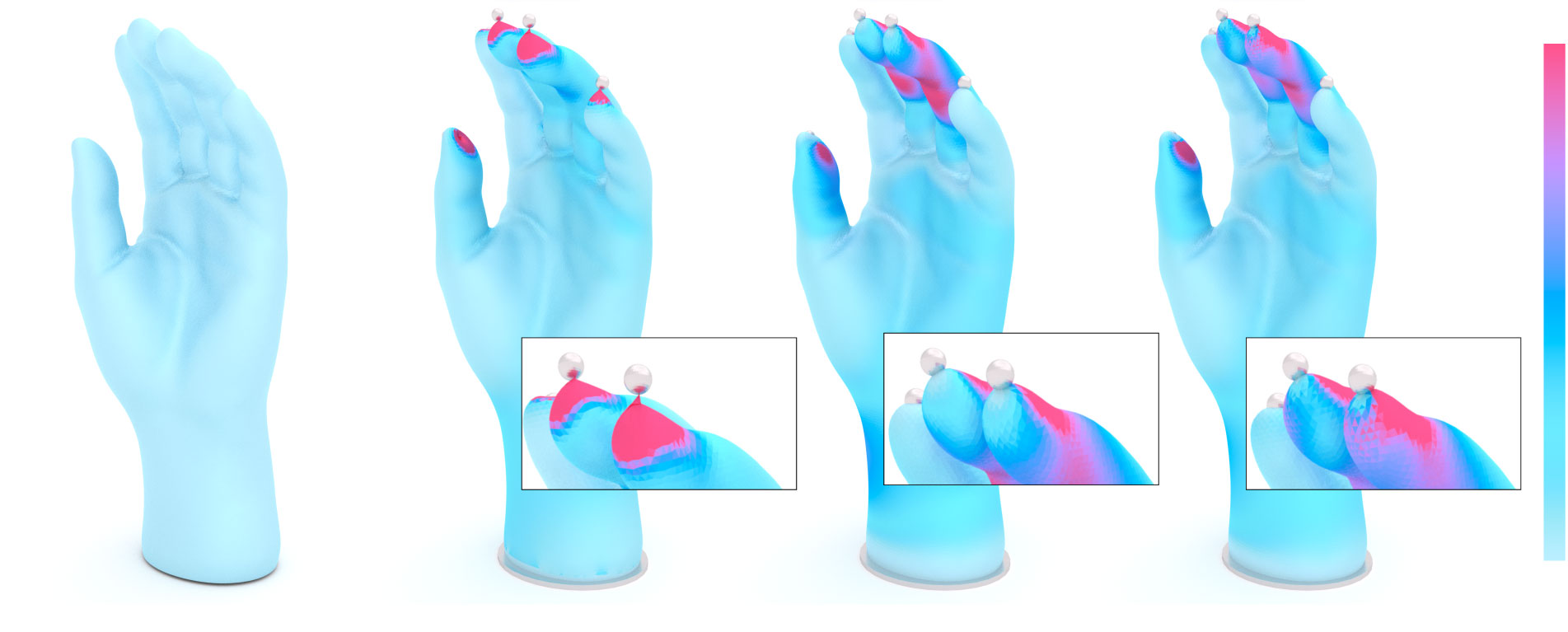
\caption{\emph{Hand}. The base of the model is fixed, and each sphere
represents a deformation handle, which was rotated. Harmonic surface deformation
is susceptible to protruding triangles and surface self-intersection near small
deformation handles. Low amounts of regularization ($\beta$) suppress these
artifacts.}
\label{fig:teaser}
\end{figure*}
\begin{abstract}

Harmonic surface deformation
is a well-known geometric modeling method
that creates plausible deformations in an interactive manner.
However, this method is susceptible
to artifacts, in particular close to the deformation handles.
These artifacts
often correlate with strong gradients of the deformation energy.
In this work,
we propose a novel formulation of harmonic surface deformation,
which incorporates a regularization of the deformation energy.
To do so, we build on and extend a recently introduced generic linear
regularization approach.
It can be expressed as a change of norm for the linear
optimization problem, \ie,
the regularization is baked into the optimization.
This minimizes the implementation complexity
and has only a small impact on runtime.
Our results show that a moderate use of regularization suppresses many
deformation artifacts common to the well-known
harmonic surface deformation method, without
introducing new artifacts.

\end{abstract}


\section{Introduction}\label{sec:introduction}

Surface deformation is an important task in geometry processing.
Deforming models involves interactive modeling sessions driven by a user, who
deforms an object by manipulating a subset of the surface vertices.
Linear deformation methods \cite{botsch2008linear} have proven effective in this
context as they often create plausible and realistic-looking deformations, while
still allowing for interactive runtimes.
Deformations are usually modeled as the minimizers of specific deformation
energies that are defined locally at each point of the domain and measure a
specific distortion property.

Harmonic surface deformation is a well-known linear gradient-domain
method introduced by Zayer \etal \cite{zayer2005harmonic} that uses a
differential surface representation to perform global mesh deformations.
Deformation constraints are smoothly distributed over the entire mesh using
harmonic functions and surface details are preserved in the reconstructed
deformations.
Furthermore, the method is parameter-free.

Linear methods such as harmonic deformations are prone to artifacts, as
linear energy terms are inadequate to accurately model the non-linear physical
forces and processes involved in a deformation.
Common deformation artifacts include flipped triangles (in the case of planar
surfaces), protruding triangles, degenerate elements, volume loss as well as
local and global shape distortion.
Many of those artifacts occur close to the deformation handles.
Figure \ref{fig:teaser} shows an example.

A number of non-linear correction methods \cite{lipman2012bounded,
aigerman2013injective, schuller2013locally, kovalsky2014singular} exists that
suppress these artifacts in planar or volumetric settings.
These methods are very powerful as they guarantee artifact-free deformations,
but they typically have a big impact on the runtime.
Most importantly, they are not applicable to deformations of surfaces.
Martinez Esturo \etal \cite{janick2014smooth} introduce an alternative linear
energy regularization method:
a quadratic regularization term is proposed that is strongly coupled to the
problem-specific deformation energy.
%
For a number of problems, this regularization yields to artifact-free results,
albeit it cannot be guaranteed.
Technically, this energy regularization requires only minor modifications to the
algorithm with little impact on the runtime.
The amount of regularization can be adjusted using a single parameter.

In this work, we apply linear energy regularization to the harmonic surface
deformation method of Zayer \etal \cite{zayer2005harmonic}.
Hereby, we follow the general ideas of Martinez Esturo \etal
\cite{janick2014smooth}.
We demonstrate that energy regularization enhances harmonic deformation results
and suppresses a variety of artifacts.
Our main contributions are:
\begin{itemize}
	\item We provide an energy-regularized formulation of harmonic surface
       	deformation.

	\item We refine the discretization of the energy differential operator
        of Martinez Esturo \etal \cite{janick2014smooth} for better estimates
        in high curvature regions.

	\item We evaluate the effectiveness of our approach.
        In particular, we demonstrate that
     	  moderate use of energy regularization improves deformation results
     	  by resolving artifacts without introducing new ones.


\end{itemize}

This paper is structured as follows:
we discuss related shape editing techniques and correction methods
(Section \ref{sec:related_work}) and review harmonic deformations as well as energy
regularization (Section \ref{sec:background}).
Then we introduce our approach to linear energy regularization for harmonic
surface deformation (Section \ref{sec:approach}).
We perform a qualitative analysis of our results (Section \ref{sec:results}), followed
by quantitative evaluation and discussion (Section \ref{sec:evaluation}).
Lastly, we present our conclusions and outlook for future work
(Section \ref{sec:conclusion}).









\begin{figure*}[ht!]%
\newlength{\horsepicwidth}%
\addtolength{\horsepicwidth}{0.18\textwidth}%
\centering%
\begin{tabular}{ccccc}%
\includegraphics[width=\horsepicwidth]{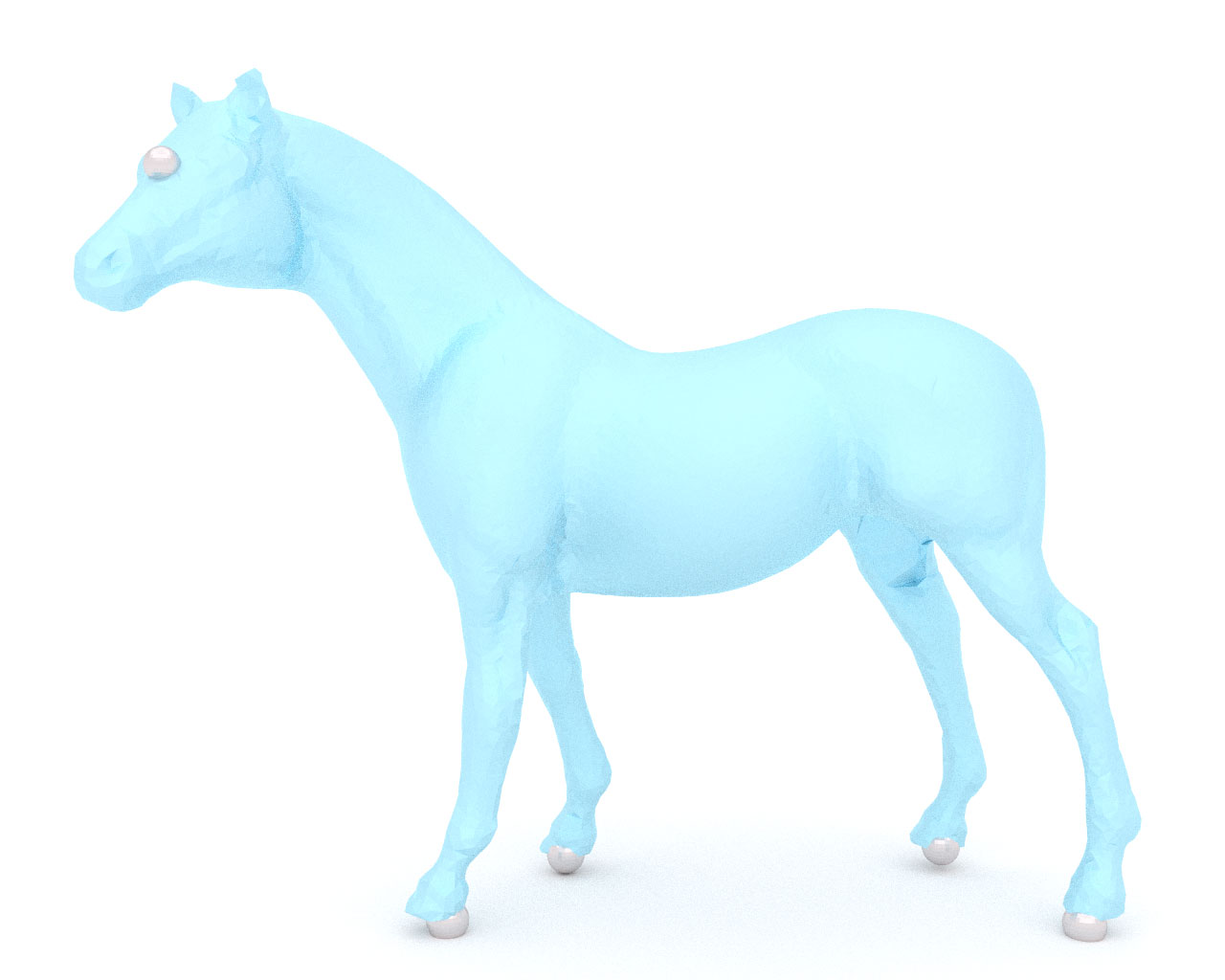} &
\includegraphics[width=\horsepicwidth]{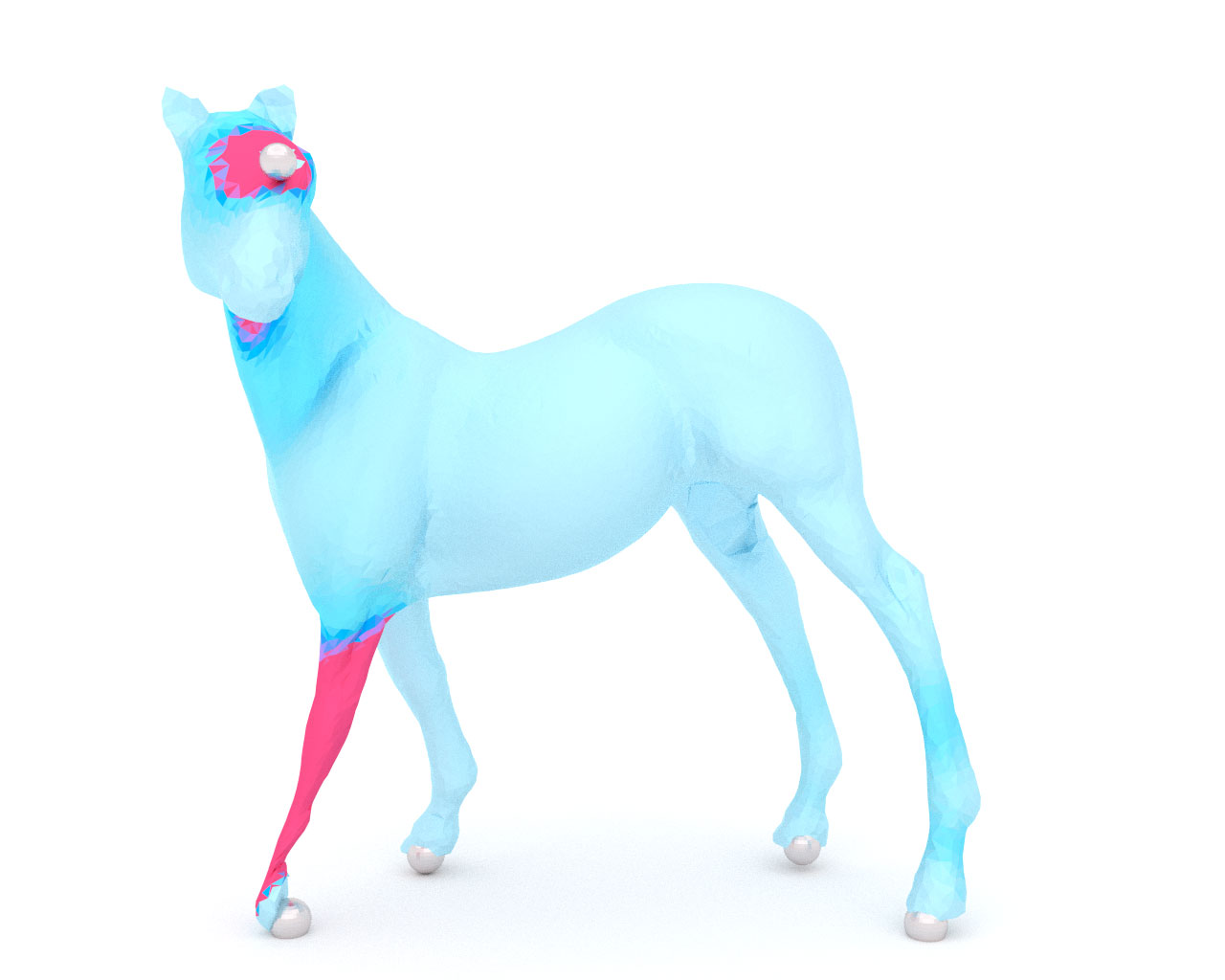} &
\includegraphics[width=\horsepicwidth]{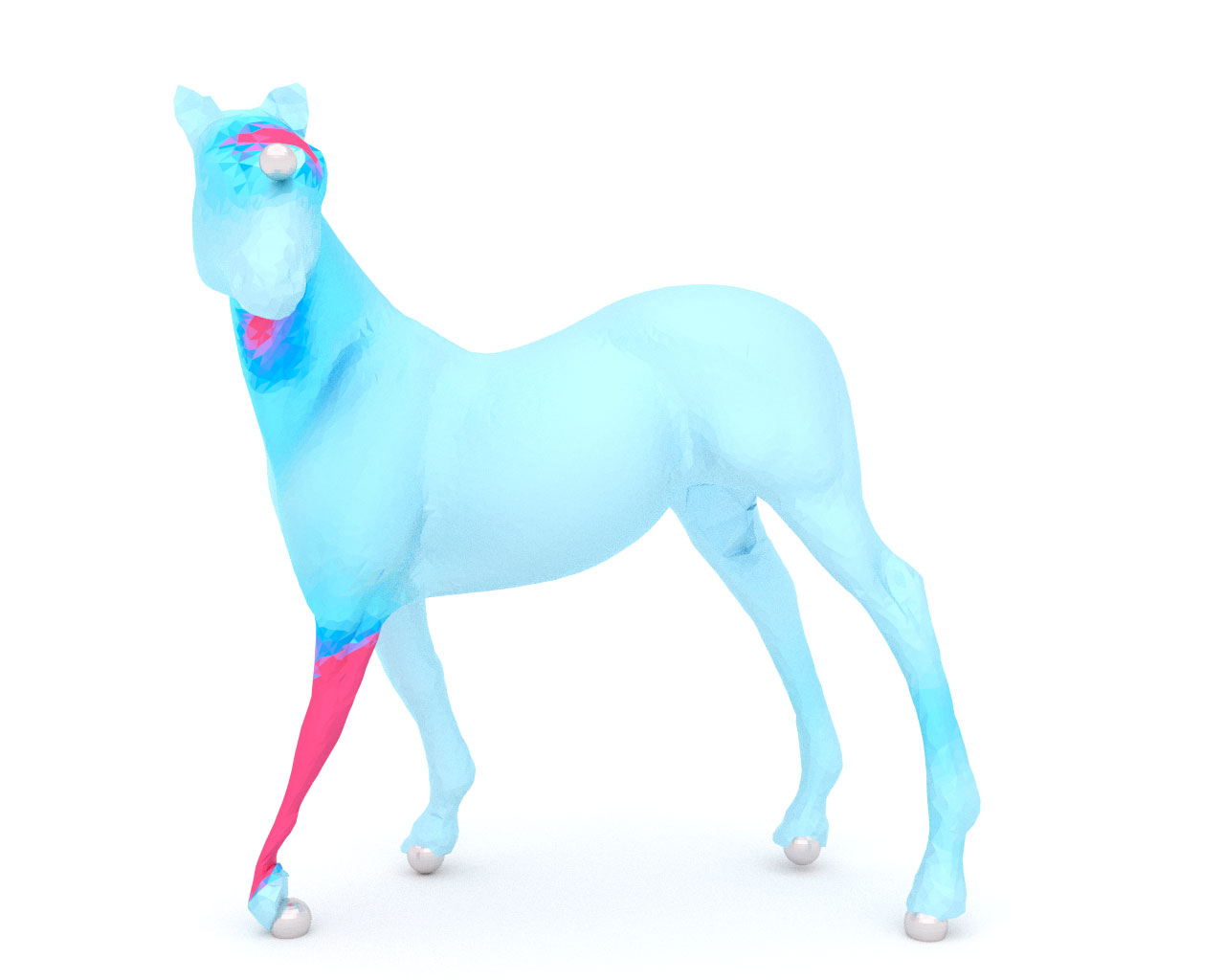} &
\includegraphics[width=\horsepicwidth]{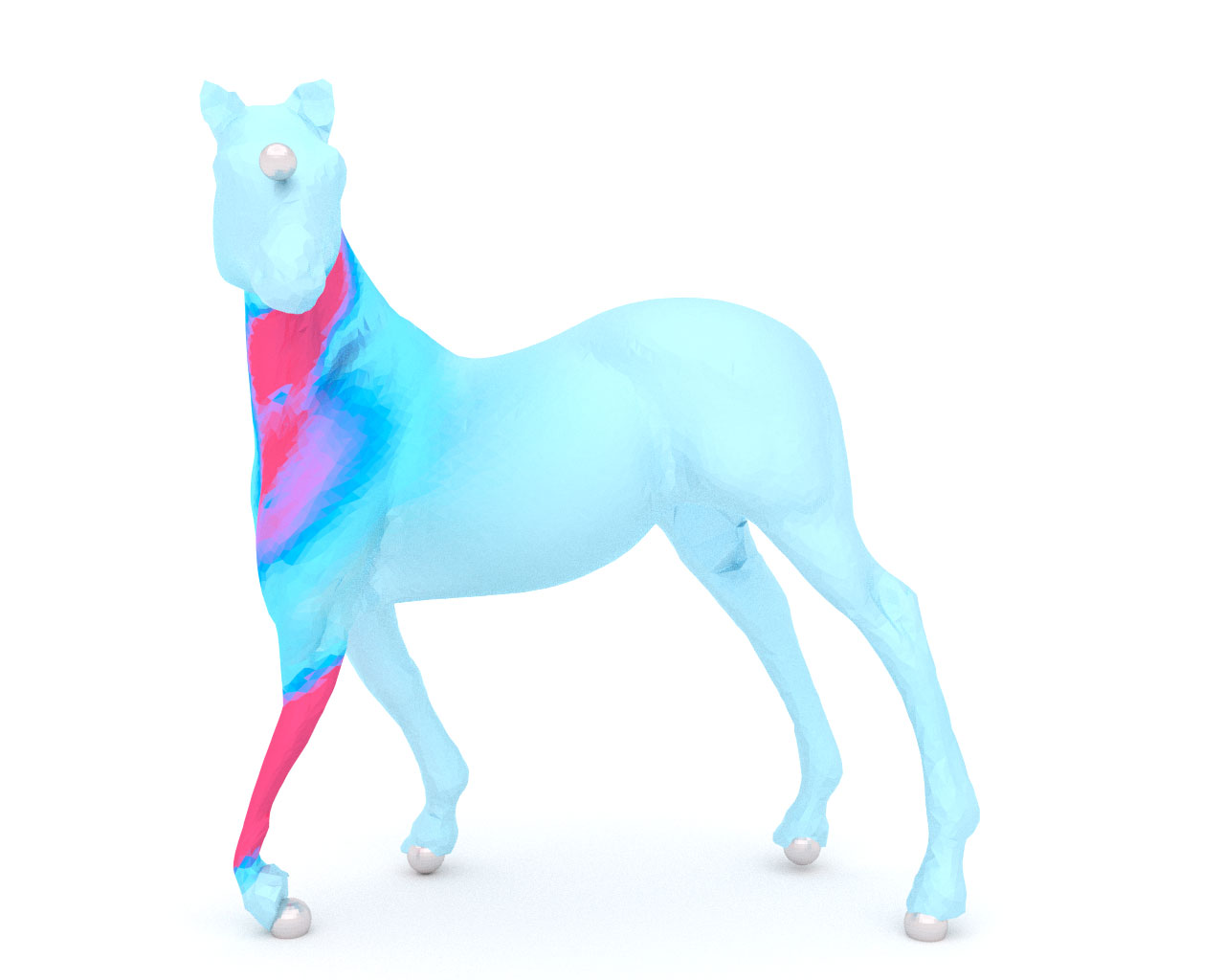}  &
\includegraphics[width=\horsepicwidth]{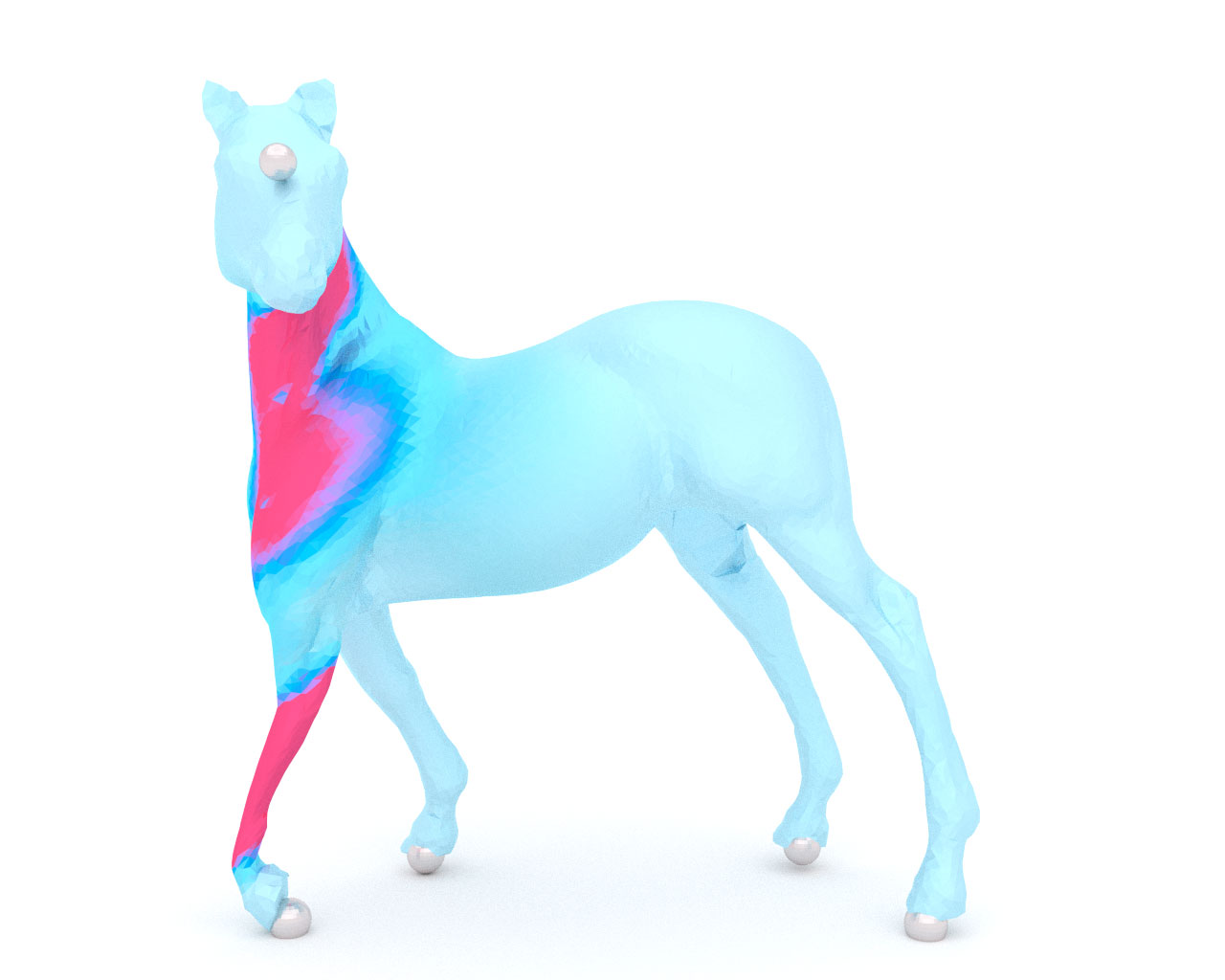}  \\
\includegraphics[width=\horsepicwidth]{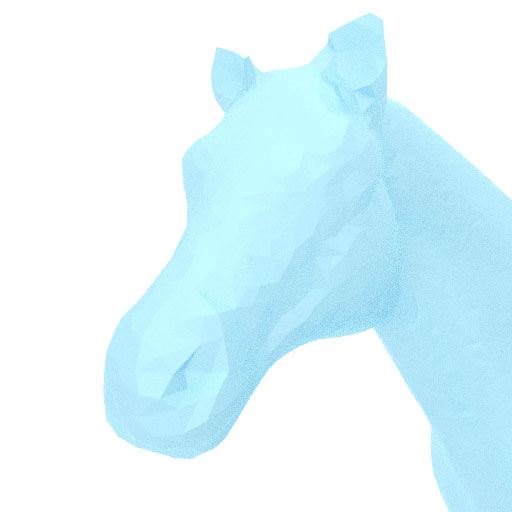} &
\includegraphics[width=\horsepicwidth]{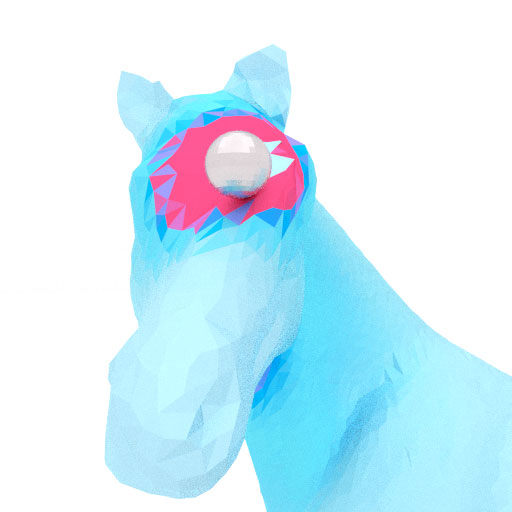} &
\includegraphics[width=\horsepicwidth]{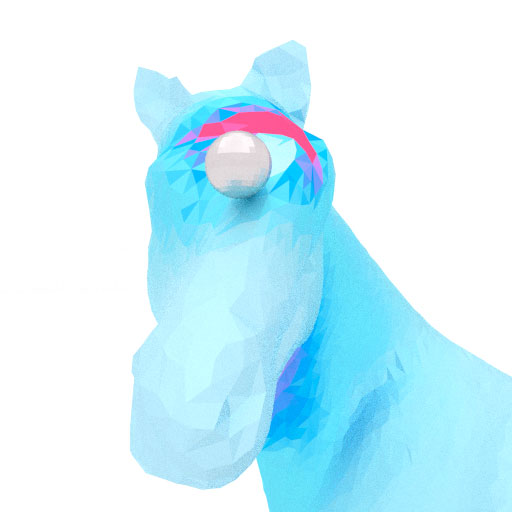} &
\includegraphics[width=\horsepicwidth]{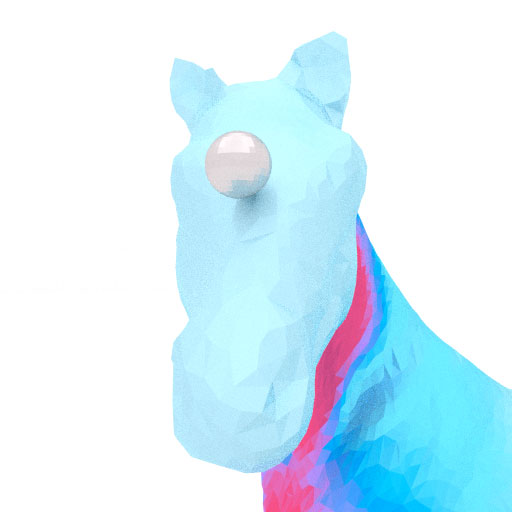}  &
\includegraphics[width=\horsepicwidth]{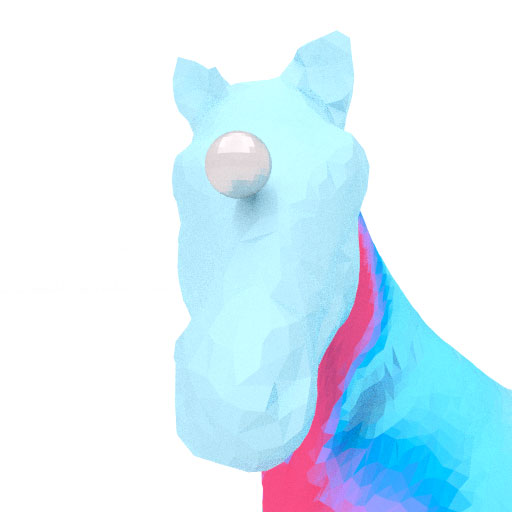}  \\[0.15cm]
\includegraphics[width=\horsepicwidth]{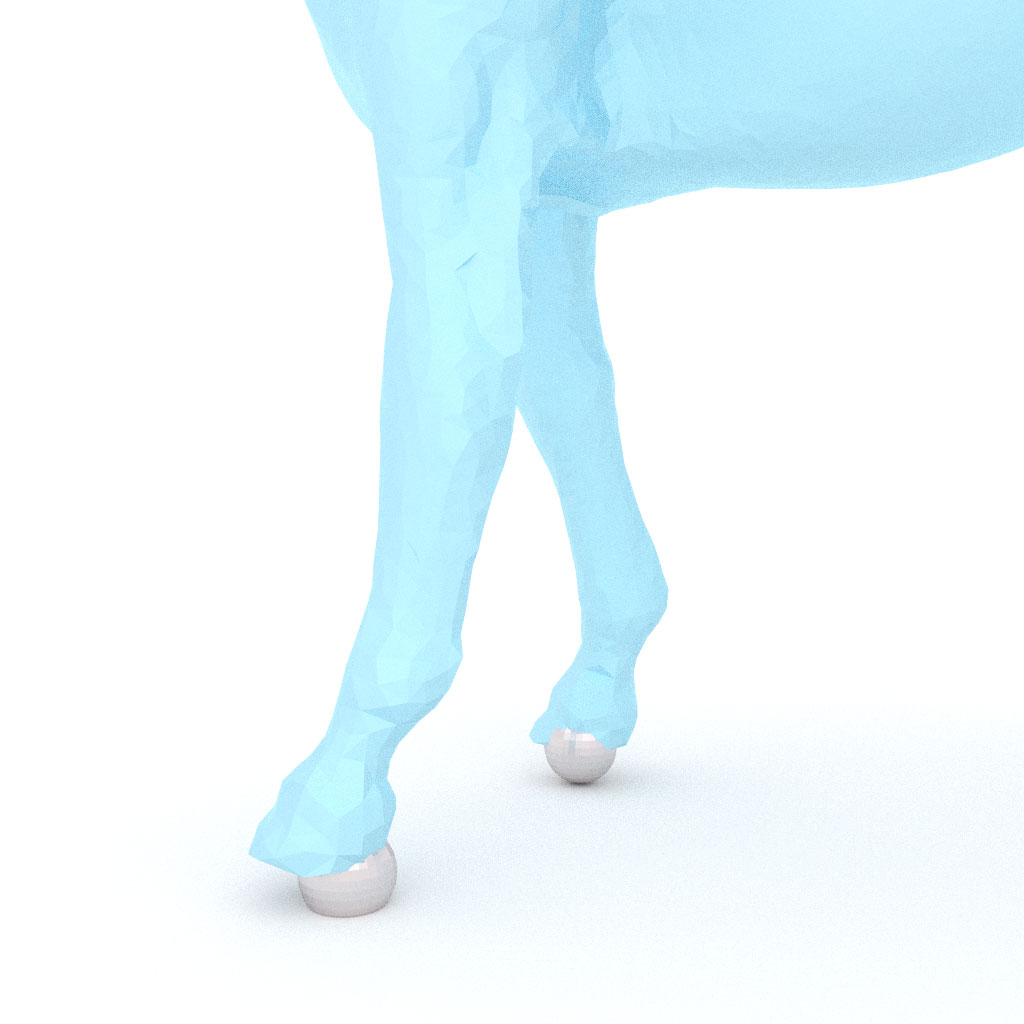} &
\includegraphics[width=\horsepicwidth]{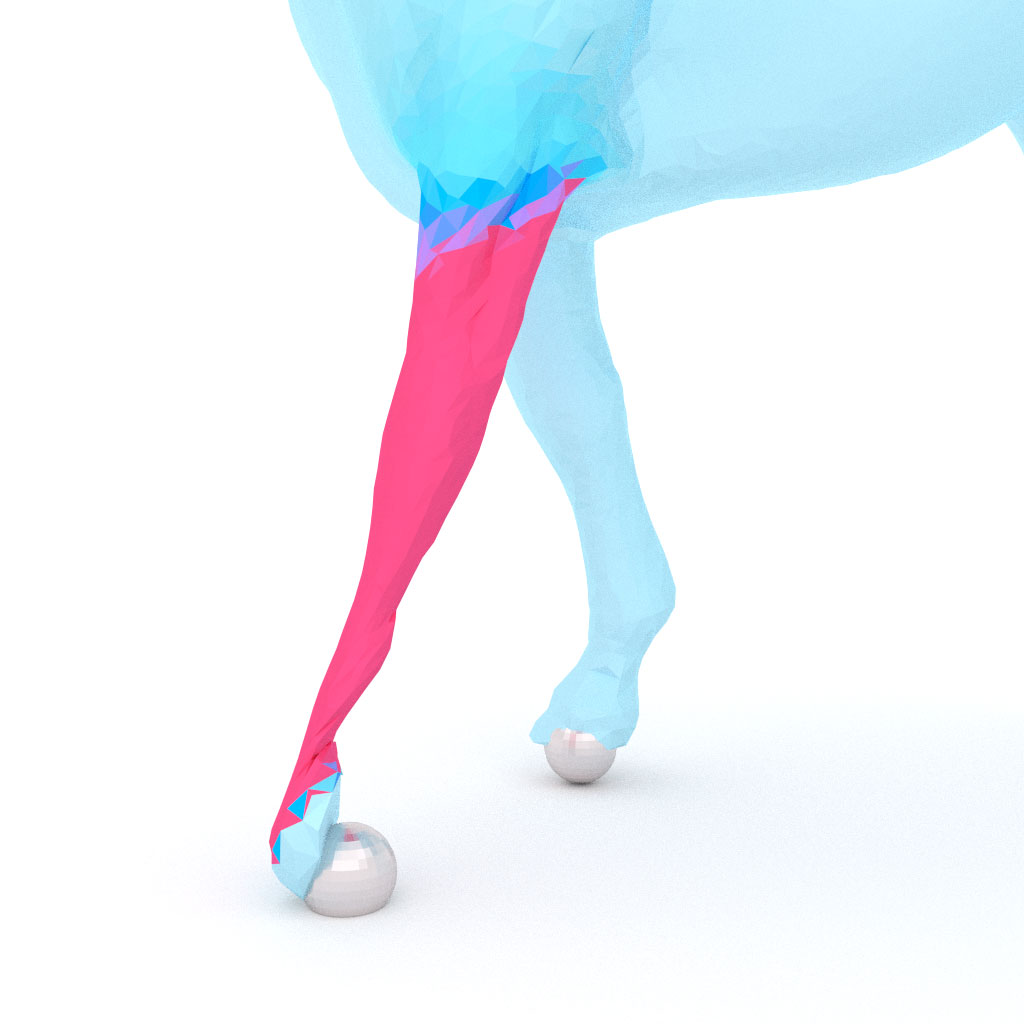} &
\includegraphics[width=\horsepicwidth]{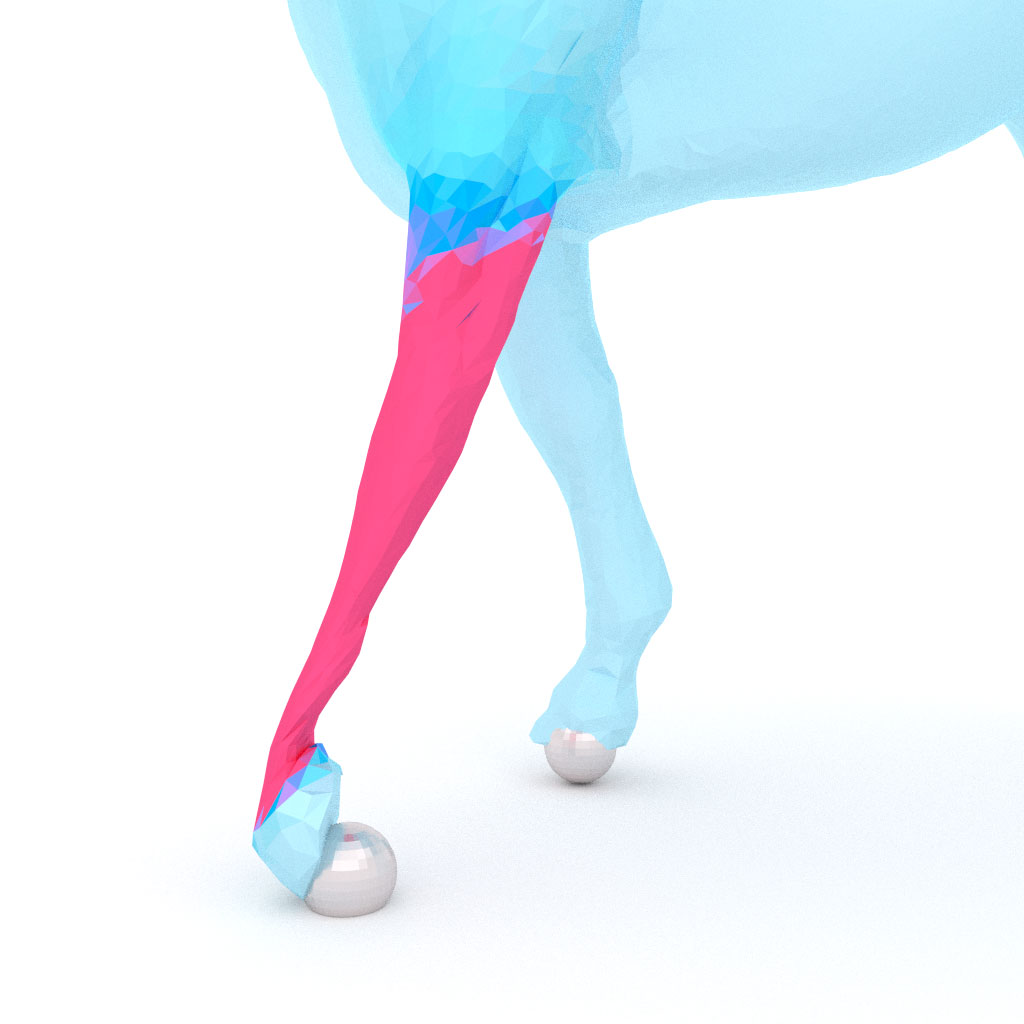} &
\includegraphics[width=\horsepicwidth]{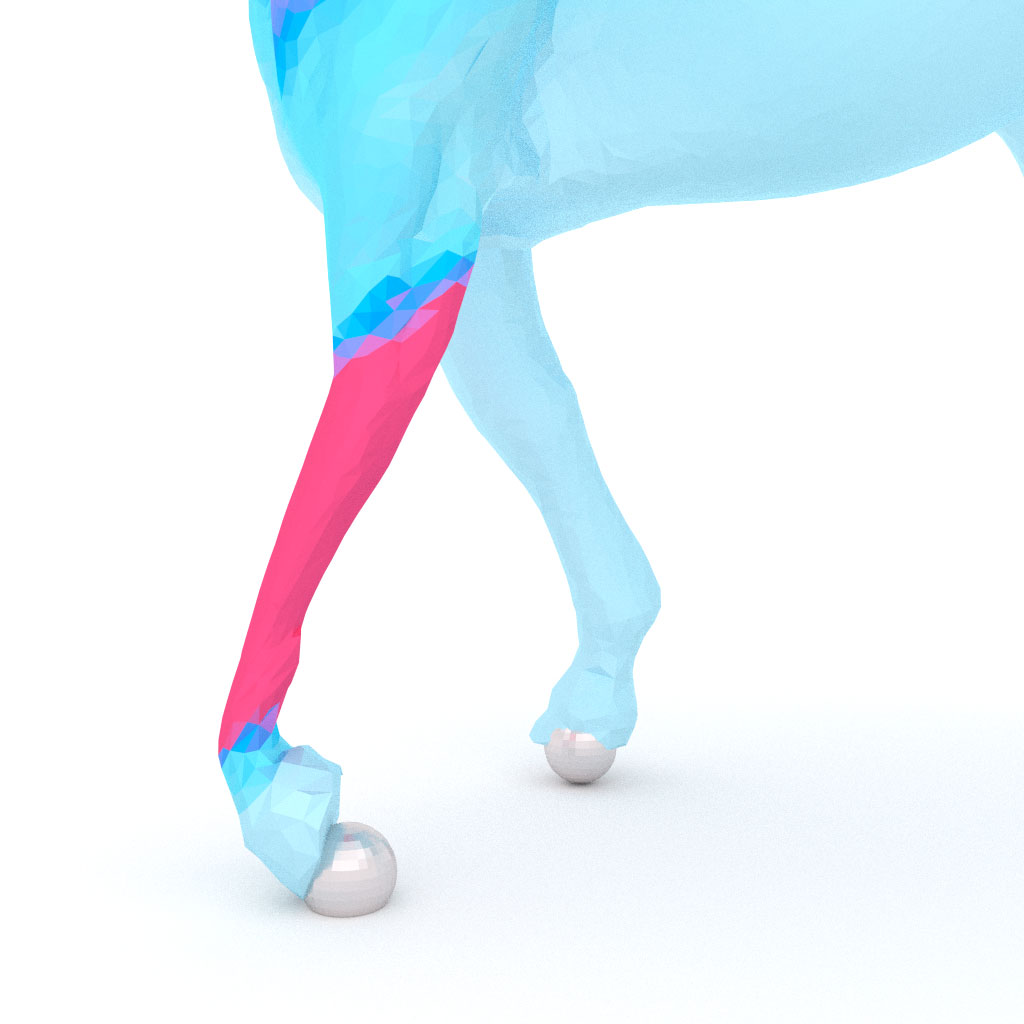}  &
\includegraphics[width=\horsepicwidth]{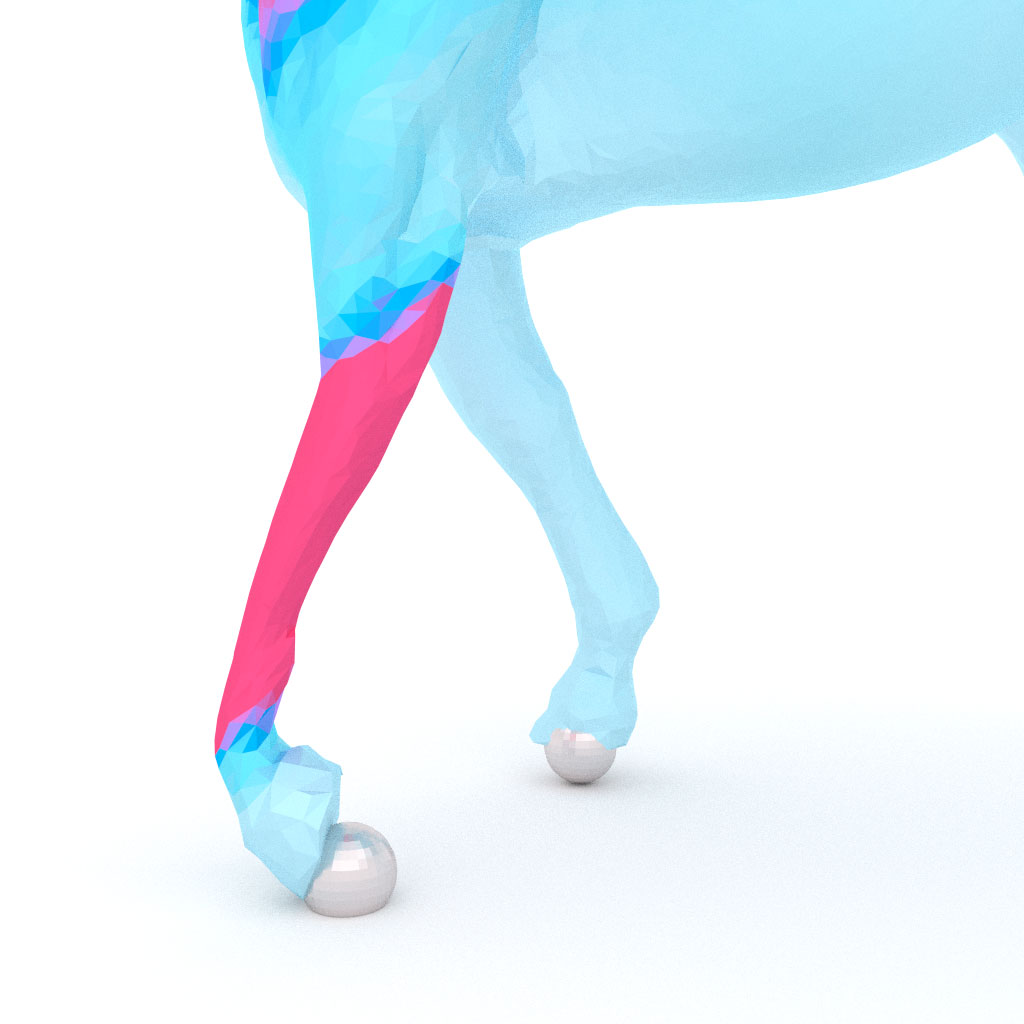}  \\
Original model & Zayer \etal \cite{zayer2005harmonic} & $\beta = 0.003$ &  $\beta = 0.1$ & $\beta = 0.2$
\end{tabular}%
\vspace{-0.35cm}%
\caption{\emph{Horse}. Each sphere corresponds to a handle region. The front
foot and the head is rotated, while the rest of the handles are kept fixed.
Harmonic surface deformation results in strong distortions close to the head,
which can be seen in the closeups (middle row), and severe volume loss at the
leg (bottom row). With increasing values of $\beta$, regularization gradually
improves volume conservation close to the foot.}%
\label{fig:horse}%
\end{figure*}%


\section{Related Work}\label{sec:related_work}

The goal of interactive surface deformation is to create meaningful deformations
while preserving surface properties such as local details and curvature. Linear
deformation methods play a major role in this area, since they provide the
interactivity and often produce plausible deformations. Most often, linear methods
represent the
surface using its differential properties
\cite{sorkine2006star}. One can distinguish these methods with
respect to their sensitivity regarding rotation and translation. Rotation
sensitive methods such as Yu \etal \cite{yu2004mesh} and Zayer \etal
\cite{zayer2005harmonic} use the gradients of affine transformations to
construct a deformation guidance field, and solve a Poisson problem for geometry
reconstruction. Since
translations introduce local changes to the tangent plane of the surface, these
methods are not suitable for shape deformations that involve large translations.
On the other hand, translation sensitive methods such as
\cite{sorkine2004laplacian} can handle large translations but not rotations.
We refer to the survey of Botsch and Sorkine \cite{botsch2008linear} for a
detailed review of linear deformation methods.

Linear techniques often cannot guarantee that the used deformations are smooth
everywhere \cite{jacobson2011bbw}. This leads to deformation artifacts such as
flipped triangles, protruding elements and volume loss due to rotations.


Artifacts can be avoided by improving the smoothness of the transformation interpolation field.
However, bi- or tri-harmonic weights
create additional local extrema in the interpolation field
that lead to unintuitive deformations results.
Jacobson \etal \cite{jacobson12a} tackle this problem
by forcing a desired topology for the interpolation field.
This requires solving a non-linear conic problem.

%

Serveral correction methods have been explored as another means to reducing
artifacts in various geometry processing tasks.
Lipman \cite{lipman2012bounded} presents a generic tool for constructing
orientation preserving (\ie, no triangle flips allowed) triangle mesh
mappings, while limiting worst-case conformal distortion. This method has
non-interactive run times and is only defined for planar meshes.
Sch\"uller \etal \cite{schuller2013locally} propose a specialized optimization
based on a barrier energy function to repress zero-area elements and
flipped triangles at interactive rates. Their iterative scheme
solves for an injective mapping to a new mesh configuration. They guarantee
inversion-free mappings of planar triangular and volumetric tetrahedral meshes.
Aigerman and Lipman \cite{aigerman2013injective} extend \cite{lipman2012bounded}
to volumetric meshes. Their algorithm takes a deformation created by common
deformation techniques and returns a similar deformation that is injective and
minimizes the distortion of the mesh volumetric elements. The method is not
interactive.
Most recently, Kovalsky \etal \cite{kovalsky2014singular} present a method based
on linear matrix inequalities for restricting the range of singular values.
It enables, \eg, bounded distortion mappings of planar or volumetric domains,
but is also computationally to expensive for interactive applications.

These correction methods guarantee that deformations are inversion-free, and
in some cases even protrusion-free.
However, they all require solving non-linear
systems, which results in loss of interactivity for moderate to large meshes.
In contrast, we follow the recent linear approach to regularization by Martinez
Esturo \etal \cite{janick2014smooth}, which has no significant impact on
runtime.
While we cannot guarantee artifacts-free deformations, our method successfully
suppresses usual deformation artifacts.
Furthermore, the correction methods mentioned above are not applicable to
surface meshes embedded in $\mathbbr^3$.
In contrast, our method is well-define for surface meshes.


\section{Background}\label{sec:background}














In this Section, we continue to review the formal details required in our work.
We consider triangulated surface meshes $\cM = (\cT, \cV, \cE)$ defined by sets
of vertices $i \in \cV$, oriented edges $\cE \subset \cV^2$, and triangles $\cT
\subset \cV^3$.
Coordinates of vertices $i \in \cV$ are denoted by $\vx_i \in \RRSet^3$.
A missing subscript either indicates a vector of stacked coefficients, \eg, $\vx
\in \RRSet^{3 \Abs{\cV}}$, the vector of stacked vertex coordinates $\vx_i$, or
a matrix of component-wise coefficients, \eg, $\mX \in \RRSet^{\Abs{\cV} \times
3}$.
For a triangle $t\in\cT$, $\mX_t \in \RRSet^{3 \times 3}$ denotes the
column-wise concatenation of the coefficients of its vertices.
Using the notations
\begin{equation}
  \invn{\vy}^2_\mN = \T{\vy}\mN\,\vy%
  \quad\text{and}\quad%
  \invn{\mY}^2_\mN = \Tr{\T{\mY}\mN\,\mY}~,
  \label{eq:norms}
\end{equation}
we denote (squared) vector and matrix norms that are induced by symmetric and
positive definite matrices $\mN$.
($\Tr{\cdot}$ denotes the trace of a matrix.)

For the piecewise linear functions on $\cM$ a discrete gradient operator $\mG
\in \RRSet^{3\ntriangles \times \nvertices}$ can be assembled from local
per-triangle gradient operators $\mG_t$:
for triangles $t = (i,j,k) \in \cT$ with normalized normals $\vn_t$, the local
gradient operators are given by
\begin{equation}
  \mG_t = \begin{bmatrix} \T{\inp{\vx_j - \vx_i}} \\
                          \T{\inp{\vx_k - \vx_i}} \\
                          \T{\vn_t}               \end{bmatrix}^{-1}
                          \begin{pmatrix} -1 & 1 & 0 \\
                                          -1 & 0 & 1 \\
                                           0 & 0 & 0 \end{pmatrix} ~,
  \label{eq:gradop_3d}
\end{equation}
see, \eg, \cite{botsch2008linear}.
Given a scalar function on $\cM$ defined by the vertex-based coefficients $\vu
\in \RRSet^{\Abs{\cV}}$, $\mG\,\vu$ is the vector of stacked and constant
per-triangle gradients.
Note that gradients computed by $\mG$ are defined in a common coordinate system.

\subsection{Harmonic Guidance for Surface Deformation}

Zayer \etal \cite{zayer2005harmonic} propose a variant of gradient-domain
deformations in which local deformation constraints are propagated using
harmonic functions.
Deformed surfaces are reconstructed from manipulated surface gradients
by minimizing the global deformation energy
\begin{equation}
  \label{eq:deformation_energy_sum}
  E(\vx) = \sum_{t \in \cT} A_t \, \invn{\mG_t \, \mX_t - \mZ_t}^2_F
\end{equation}
subject so suitable boundary constraints.
Here, $A_t$ denotes the area of triangle $t$, $\invn{\mM}^2_F = \Tr{\T{\mM}\mM}$
is the (squared) Frobenius norm of $\mM$, and $\mZ_t \in \RRSet^{3 \times 3}$
are prescribed component-wise \emph{guidance gradients} that are constant per
triangle.
Dense guidance gradients are computed from user-specified transformations
associated to a set of handle regions.
In \cite{zayer2005harmonic}, harmonic functions $h(\vx)$ given by solutions of
the Poisson equation $\mL \, \vh = \vNull$ are used for the global propagation
of the sparse set of given transformations.
Here, $\mL = \T{\mG} \mA \, \mG$ is a discretization of the Laplace-Beltrami
operator \cite{botsch2008linear}, in which $\mA$ is a diagonal matrix of
replicated triangle areas.
Please, see Zayer \etal \cite{zayer2005harmonic} for further details
on the quaternion-based propagation of transformations using harmonic functions.
As minimizers of \eqref{eq:deformation_energy_sum} are characterized by the
same differential operator, in practice a single factorization of $\mL$ can
be used to perform both transformation propagation and energy minimization.

\paragraph*{Drawbacks.} Linear deformation methods are susceptible to various
artifacts.
This is because the physical energies involved in deformations are non-linear by
nature, and are only approximated by linear methods \cite{botsch2008linear}.
Specifically, harmonic surface deformation is susceptible various deformation
artifacts, which are in particular related to the size of the handle regions.
Small deformation handles are likely to cause protruding and intruding
triangles.
Large deformation handles cause local shape distortion on the boundary between
the constrained and free mesh areas.
Other artifacts include volume loss close to deformation handles when the
deformation includes large rotations and local surface intersection.
It was observed \cite{janick2014smooth} that these artifacts correlate with
large spatial variation in the optimized energy on the domain.

\subsection{Linear Energy Regularization}\label{sec:linear_energy_ref}
Martinez Esturo \etal \cite{janick2014smooth} propose a generic linear
energy regularization scheme that suppresses geometric artifacts in a number
of different applications.
It is applicable to regularize problem-specific squared energies of the general
form
\begin{equation}\label{eq:discrete_energy}
	E_\cP(\vu) = \invn{\mE \, \vu - \vc}_\mA^2
\end{equation}
over $d$-dimensional piecewise linear functions defined by vertex-based
coefficients $\vu \in \RRSet^{d\Abs{\cV}}$.
Here, $\mE \in \RRSet^{n\Abs{\cT} \times d\Abs{\cV}}$ is a problem-specific
linear energy operator that maps unknown functions $\vu$ to triangle-constant
\emph{local energies} of dimension $n$, $\vc$ are problem-specific energy
constants, and $\mA$ is a diagonal matrix of replicated triangle areas that
performs domain-wide integration of triangle-constant quantities.
Energies $E_\cP$ are regularized by introducing a regularization term
\begin{equation}\label{eq:discrete_regularization}
	E_\cR(\vu) = \invn{\mD \, \inp{\mE \, \vu - \vc}}_\mB^2
\end{equation}
that measures squared \emph{variations} of local energies.
For piecewise constant local energies, pointwise energy variations are estimated
by the sparse differential operator $\mD$.
\setlength\intextsep{-1pt}
\begin{wrapfigure}[5]{r}[0cm]{0cm}
  \hspace{-23pt}
  \def\svgwidth{0.245\linewidth} \small 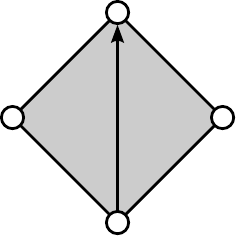
\end{wrapfigure}
For each pair of neighboring triangles, it can be discretized along all internal
edges $e \in \cE_i$ from the set of non-boundary edges $\cE_i \subseteq \cE$:
let $l(e)$ and $r(e)$ denote the left and right triangle at $e$, respectively.
Then, for scalar local energies ($n=1$), the nonzero coefficients of $\mD$
are given by
$D_{et}=\left\{\def\arraystretch{1.0}\begin{tabular}{@{}l@{\;\;}l@{}}
  $1$  & if $l(e)=t$ \\
  $-1$ & if $r(e)=t$
\end{tabular}\right.$, for all internal edges $e\in\cE_i$ and triangles
$t\in\cT$.
For vector-valued local energies ($n>1$), the differential operator is given by
a component-wise replication, which can be expressed as $\mD \otimes \mI_n$ using
the Kronecker product $\otimes$ and the $n \times n$ identity matrix $\mI_n$.
The constant estimates of pointwise local energy variations are integrated
using the diagonal matrix $\mB$ of replicated internal edge lengths.

The total regularized energy is given by a weighted combination of both terms
\begin{equation}\label{eq:discrete_total}
	E_\beta(\vu) = (1-\beta) \, E_\cP(\vu) + \beta \, E_\cR(\vu)
	= \invn{\mE \, \vu - \vc}^2_{\mW_\beta}
\end{equation}
that can be expressed compactly using the $\beta$-weighted norm
\begin{equation}\label{eq:discrete_norm}
	\mW_\beta = (1-\beta) \, \mA + \beta \, \T{\mD}\mB\,\mD ~.
\end{equation}
The amount of regularization is steered by $\beta \in [0,1)$.
Note that this formulation of energy regularization is also valid for energies
in the components of the unknown functions $\vu$, which are then given by
$E_\beta(\mU) = \invn{\mE \, \mU - \mC}^2_{\mW_\beta}$.
Please, see Martinez Esturo \etal \cite{janick2014smooth} for further details
and applications on this energy regularization scheme.


\section{Enhancing Harmonic Surface Regularization}
\label{sec:approach}

We continue to show that the concept of energy regularization is
applicable to harmonic surface deformation in a straightforward way.
For this, we rewrite the component-wise deformation energy
\eqref{eq:deformation_energy_sum} to the equivalent formulation
\begin{equation}
  \label{eq:deformation_energy}
  E(\mX)  =  \invn{\mG \, \mX - \mZ}^2_\mA
\end{equation}
using the global gradient operator $\mG$, the matrix $\mZ$ of all stacked
prescribed gradients, and diagonal matrix $\mA$ of replicated triangle areas.
Note that the local energies correspond to the to the summed terms of
\eqref{eq:deformation_energy_sum}.
Comparing our problem-specific energy \eqref{eq:deformation_energy} and the
regularizable generic energy \eqref{eq:discrete_energy}, we obtain the
correspondences that the generic energy operator $\mE$ is given by the gradient
operator $\mG$, the constant energy term $\mC$ is given by the gradient field
$\mZ$, and the dimension of the local energies is $n=3$.
Hence, the energy-regularized version of the deformation energy
\eqref{eq:deformation_energy} is given by
\begin{equation}
  \label{eq:regularized_energy}
  E_\beta(\mX)  =  \invn{\mG \, \mX - \mZ}^2_{\mW_\beta}
\end{equation}
by simply applying the weighted norm $\mW_\beta$ for energy integration
and smoothness estimation.
Technically, to apply regularization, this substitution of $\mA$ for
$\mW_\beta$ allows for a straightforward implementation.
Specifically, the remainder of the original harmonic surface deformation
approach is unchanged, in particular the harmonic function-based transformation
propagation for the guidance gradients $\mZ$.

Our experiments demonstrate that this simple energy modification suppresses a
variety of deformation artifacts of the original energy formulation (see
Section \ref{sec:results}).
Still, the original discretization of the energy differential operator $\mD$
is defined independently of the local surface curvature, which leads to
poor estimates of energy variation in high curvature regions.
We continue to provide a refined differential operator discretization that is
based on surface curvature and yields better estimates of energy variation.

\paragraph*{Curvature-based Energy Differential Operator.}
Martinez Esturo \etal \cite{janick2014smooth} estimate the variation of local
energies by finite differences of their respective local energy residuals
(Section \ref{sec:linear_energy_ref}).
In our application of gradient-domain deformations, the energy residual $\mE_t$
of a triangle $t$ is given by the deviation between the
deformed mesh gradients and the gradients of the guidance field:
\begin{equation}
\label{eq:resudial}
\mE_t = \mG_t\,\mX_t - \mZ_t ~.
\end{equation}
%
The (squared) energy variation between two neighboring triangles $t_1$ and $t_2$
is now simply given by $\invn{\mE_{t_1} - \mE_{t_2}}^2_F$.

This estimation works well if both triangles are aligned and therefore also
the residuals in the columns of $\mE_{t_1}$ and $\mE_{t_2}$ live in the same tangent space, \eg, for
planar meshes.
However, if the tangent spaces are not aligned, this estimation is likely to
break.
The effect is particularly noticeable at sharp edges of a curved surface, where
areas of low energy values along the edge are surrounded by higher energy
values.
This results in a less smooth energy distribution across edges of high
curvature.
%


We adjust the discretization of the operator $\mD$ to also be applicable to
high curvature regions by locally compensating for curvature.
Similar local alignments of tangent spaces is used, \eg, by Crane \etal
\cite{crane2010connections}, for the computation of connections relating
neighboring tangent spaces.
\begin{wrapfigure}[6]{r}[0cm]{0cm}
  \hspace{-23pt}
  \def\svgwidth{0.25\linewidth} \tiny 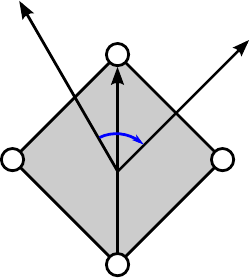
\end{wrapfigure}
For the $n=3$ dimensional local energy residuals \eqref{eq:resudial}, our refined
energy differential operator $\mD^R$ is a $\ninternaledges\times\ntriangles$
matrix of $3\times3$ block matrices.
For each internal edge $e\in\cE_i$, we use the rotation $\mR_e$ that aligns the
normals of its left and right triangles $\vn_r, \vn_l$ such that $\vn_r = \mR_e
\vn_l$.
Then, the nonzero blocks of $\mD^R$ are given by
%
$D^R_{et}=\left\{\def\arraystretch{1.0}\begin{tabular}{@{}l@{\;\;}l@{}}
  $\mR_e$  & if $l(e)=t$ \\
  $-\mI_{3}$ & if $r(e)=t$
\end{tabular}\right.$
for all internal edges $e\in\cE_i$ and triangles $t\in\cT$.
This way, $\mD^R$ compensates for local curvatures:
residuals that are similar relative to their respective tangent spaces are not
estimated to be different anymore.
Note that this operator refinement simplifies to the original formulation for
planar meshes.
It reduces the sparseness of the resulting linear system only by a constant
factor, which does not impair resulting runtimes.
However, it is only applicable to this particular case of $n=3$ dimensional
local energy residuals.
For different values of $n$ the discretization should therefore fall back to the
$\mD$ operator of \cite{janick2014smooth}.

\paragraph*{Implementation.}
The operators $\mG$, $\mD^R$, $\mA$, and $\mB$ as well as $\mL$ are assembled
once when the surface mesh is loaded.
For each deformation of the model the guidance field $\mZ$ is computed and the
corresponding normal equations
\begin{equation}\label{eq:pois_mw_beta}
  \mG^{\transpose} \mW_{\beta} \, \mG \, \mX = \mG^{\transpose} \mW_{\beta}
  \, \mZ
\end{equation}
of \eqref{eq:regularized_energy} are solved for the coordinates $\mX$ of the
deformed mesh.
The norm $\mW_{\beta}$ is assembled for a given $\beta$ value using the refined
differential operator $\mD^R$.
After elimination of positional hard boundary constraints, the system
\eqref{eq:pois_mw_beta} is symmetric positive definite and it is solved using a
Cholesky factorization with fill-in reducing reordering \cite{eigenweb}.
Similar to \cite{zayer2005harmonic}, this factorization can be reused for
different guidance fields $\mZ$ as long as the handle configuration and $\beta$
values are unchanged.


\section{Results}\label{sec:results}
%
%
%

We evaluate our approach qualitatively on a number of different models.
The models presented have $3-36.6k$ vertices.
Deformations are performed using single and multiple handles, varying sizes of
the handle regions and the regularization weight $\beta$.

\paragraph*{Energy Visualization.}
We use color to visualize the squared magnitude of local triangle-constant
energies $\invn{\mE_t}^2_F$.
Energies are linearly mapped to the color space (shown in Figure
\ref{fig:teaser}) by setting the maximum of the color interval to correspond to
the $95$th percentile of the energy values.
The top $5\%$ of the energy values is clipped to allow visualization of the
variation of lower energies with higher contrast.

\begin{figure}
	\newlength{\cactussidepicwidth}
	\addtolength{\cactussidepicwidth}{0.22\columnwidth}

	\centering
	\begin{tabular}{cccc}
	\includegraphics[width=\cactussidepicwidth]{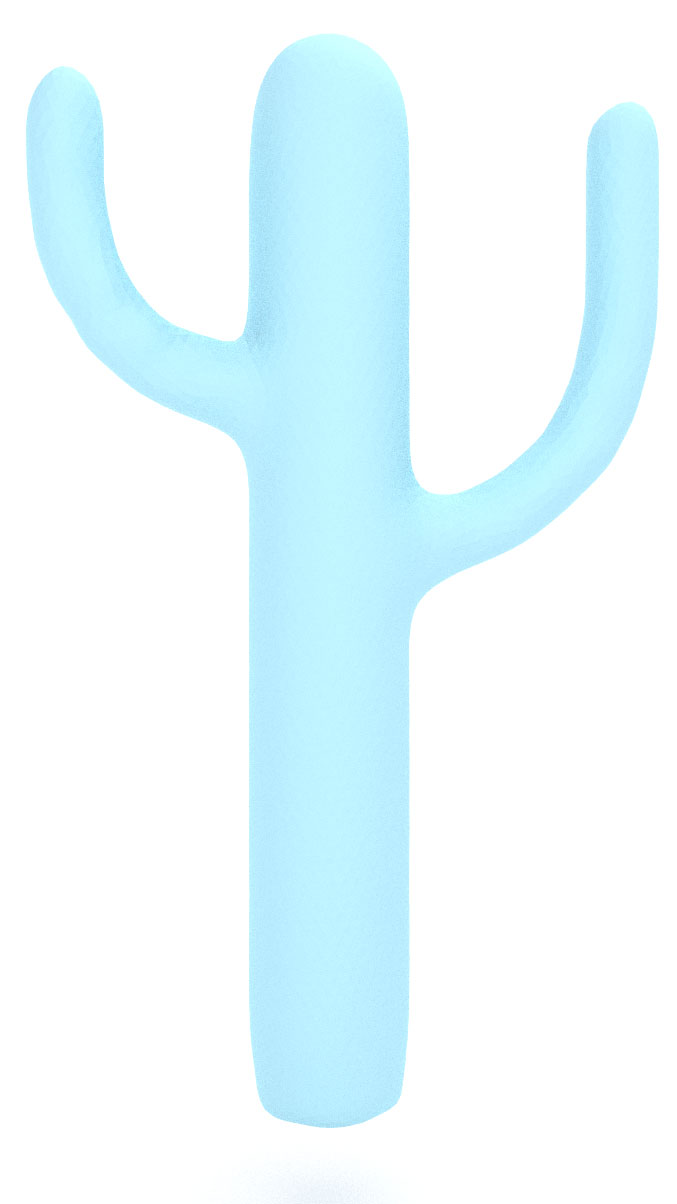} &
	\includegraphics[width=\cactussidepicwidth]{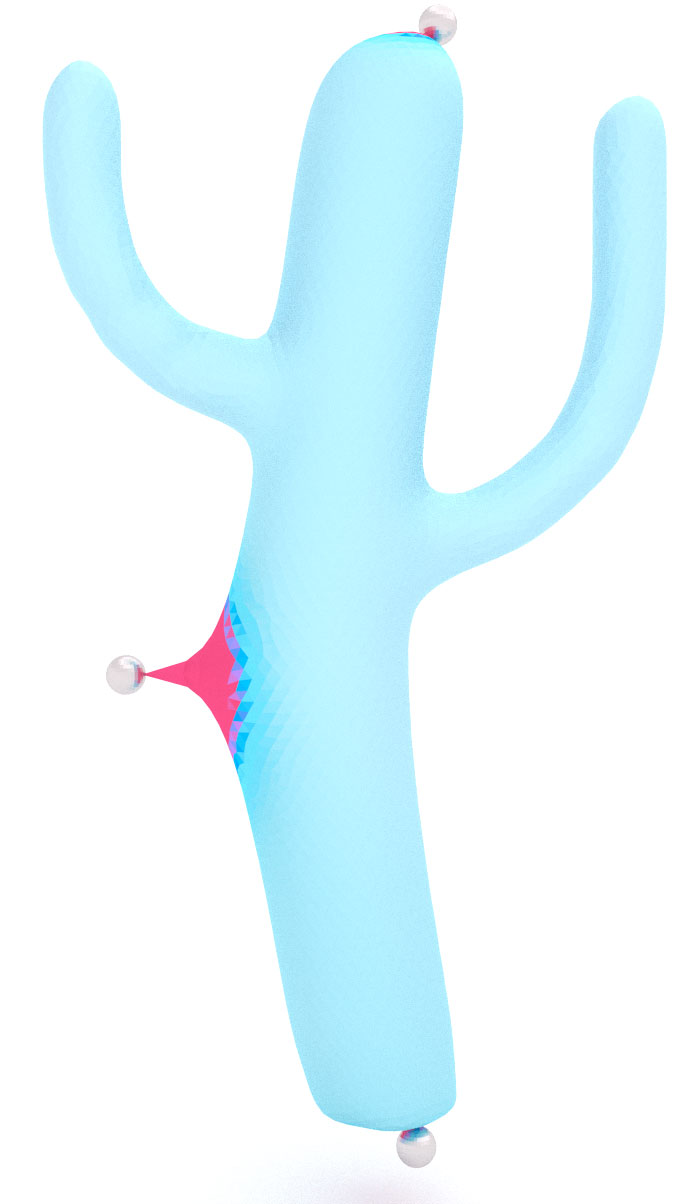} &
	\includegraphics[width=\cactussidepicwidth]{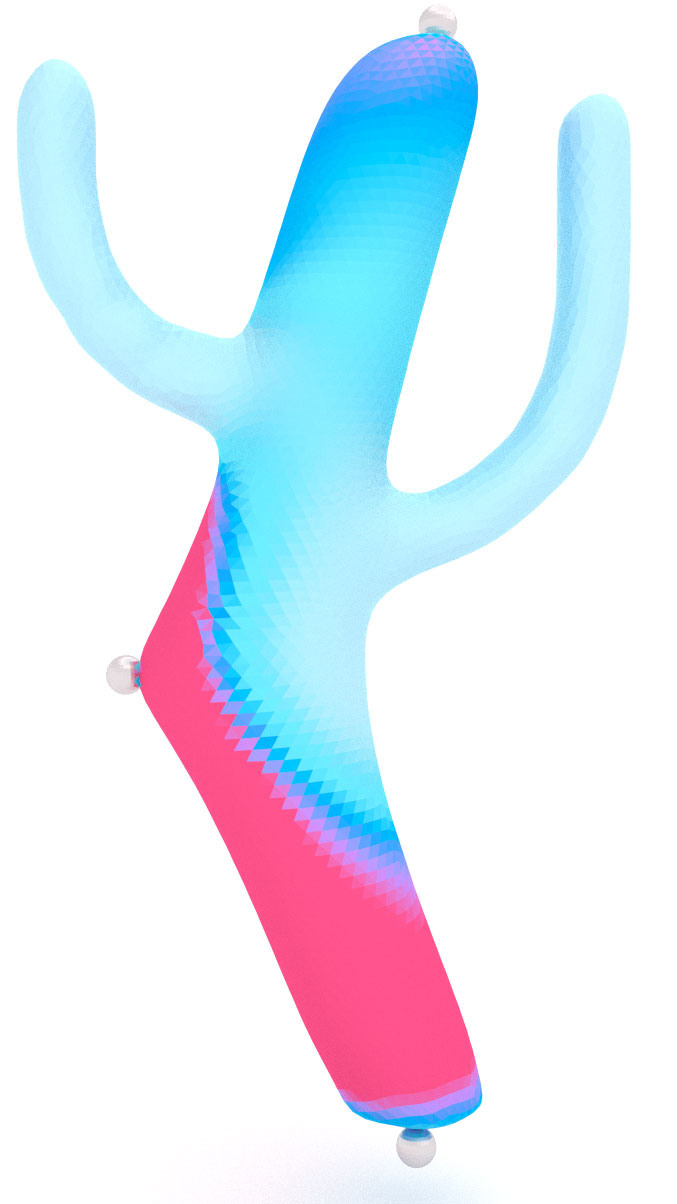} &
	\includegraphics[width=\cactussidepicwidth]{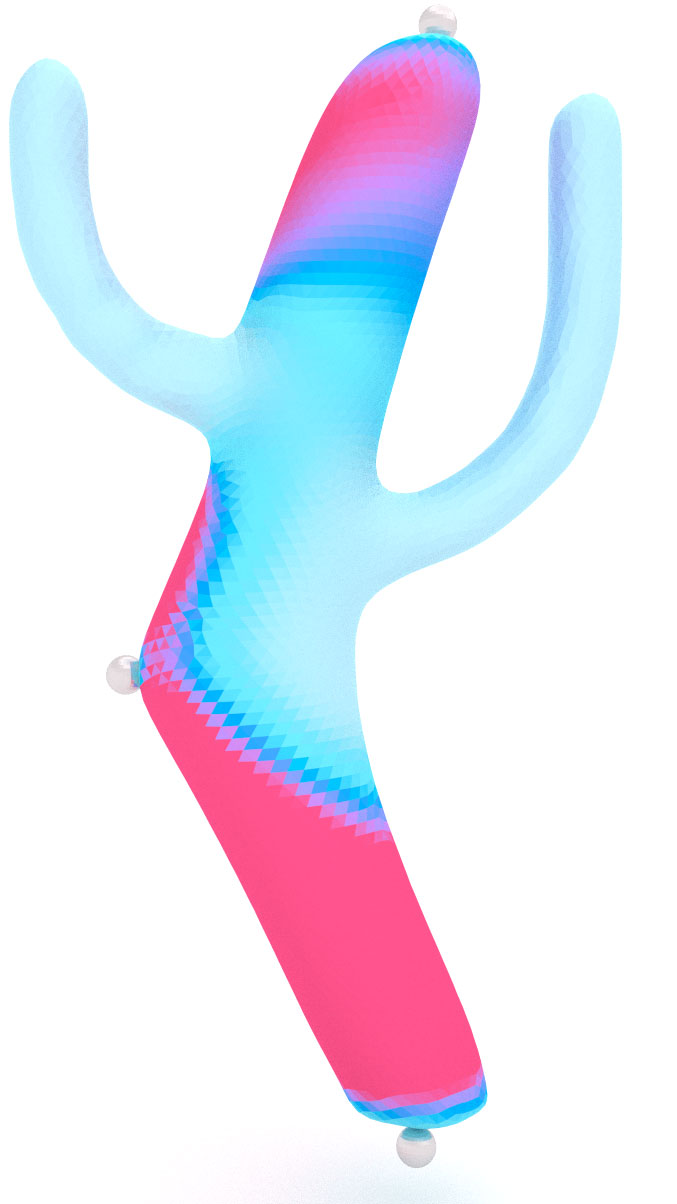}  \\
	Original & Zayer \etal  &  $\beta = 0.1$ & $\beta = 0.2$ \\
	model & \cite{zayer2005harmonic} \vspace{-0.35cm}
	\end{tabular}
	\caption{\emph{Cactus 2}. Single vertices at the bottom and
	at the top of the cactus are fixed, and a single vertex is pulled away from
	the cactus body. Regularization results in deformations free of
	protruding triangles.}
	\label{fig:cactus_side}
\end{figure}

\paragraph*{Small Deformation Handles.}
For small handle regions, harmonic surface deformation tends to create
artifacts such as protruding triangles and local surface intersections close
to the handles.
We show two examples of these artifacts:
the \emph{Hand} deformation (Figure \ref{fig:teaser}) is created by fixing the base of
the model, and a single vertex on each finger acts as the deformation handle.
All handles are rotated inwards, resulting in protruding triangles and local
surface intersections near the deformation handles.
Regularization ($\beta > 0$) suppresses these artifacts.
Similar artifacts to protruding triangles, albeit of smaller magnitude, occur if
handle regions include up to several dozen vertices.
In the \emph{Cactus 2} example shown in Figure \ref{fig:cactus_side}, single vertices
on the base and the top are fixed, and a single vertex on the side of the cactus
is used as a deformation handle.
The deformation suffers from large protruding triangles.
Introducing regularization corrects this artifact.
For both deformations, even low amounts of regularization result in a
smoother energy distribution, affecting more triangles in the mesh.
This way the optimization favors more global changes to the mesh instead of
local concentrations of energies, which result in the observed local artifacts.
%

\begin{figure}[t]
	\centering
	\begin{tabular}{ccc}

	\includegraphics[width=0.3\columnwidth]{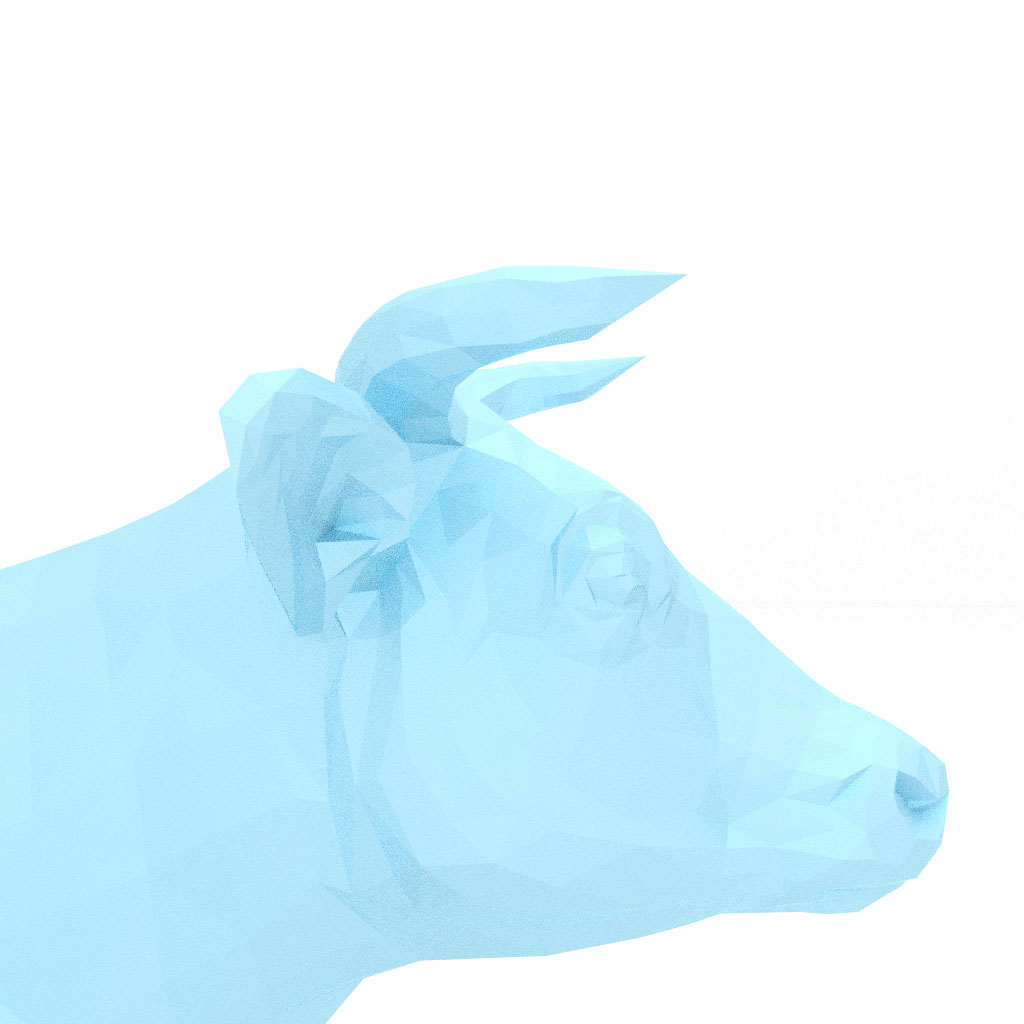} &
	\includegraphics[width=0.3\columnwidth]{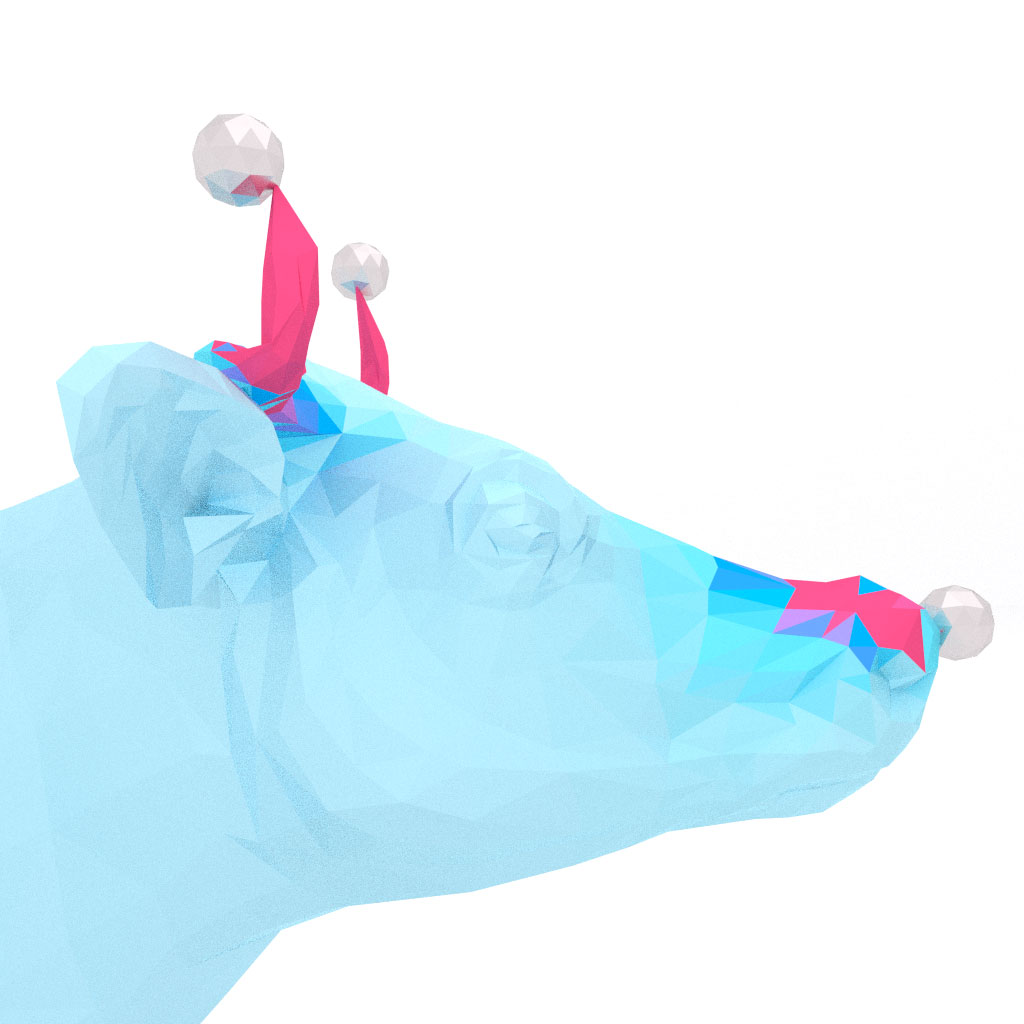} &
	\includegraphics[width=0.3\columnwidth]{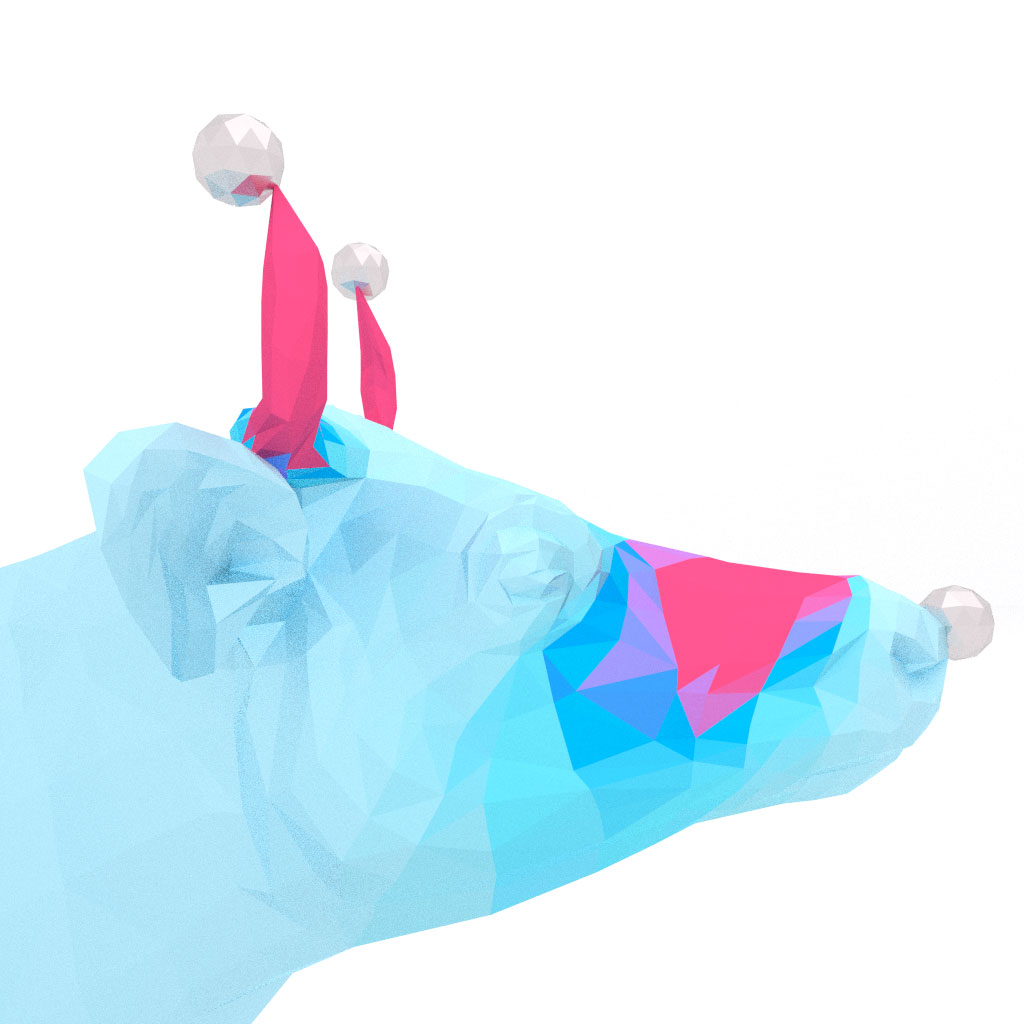} \\
	Original model & Zayer \etal  & $\beta = 0.25$ \\
	& \cite{zayer2005harmonic} \vspace{-0.35cm}
	\end{tabular}
	\caption{\emph{Cow}. The handle on face is fixed, and the horns are rotated upwards.
	Regularization helps preserve the volume near the base of the horns.}
	\label{fig:cow_face}
\end{figure}

\begin{figure*}
	\centering
	\begin{tabular}{cccc}

\includegraphics[width=0.22\textwidth]{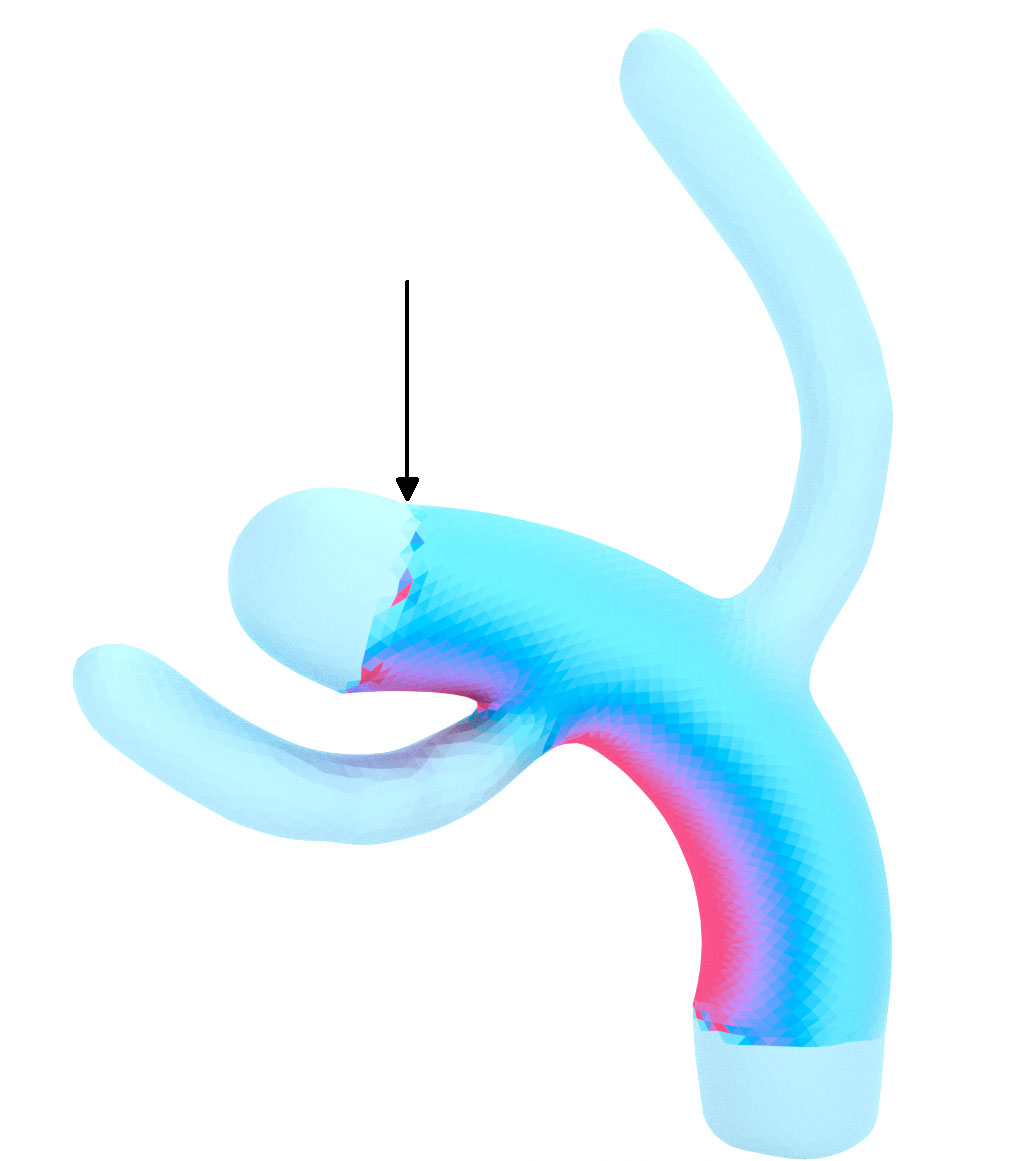} &
\includegraphics[width=0.22\textwidth]{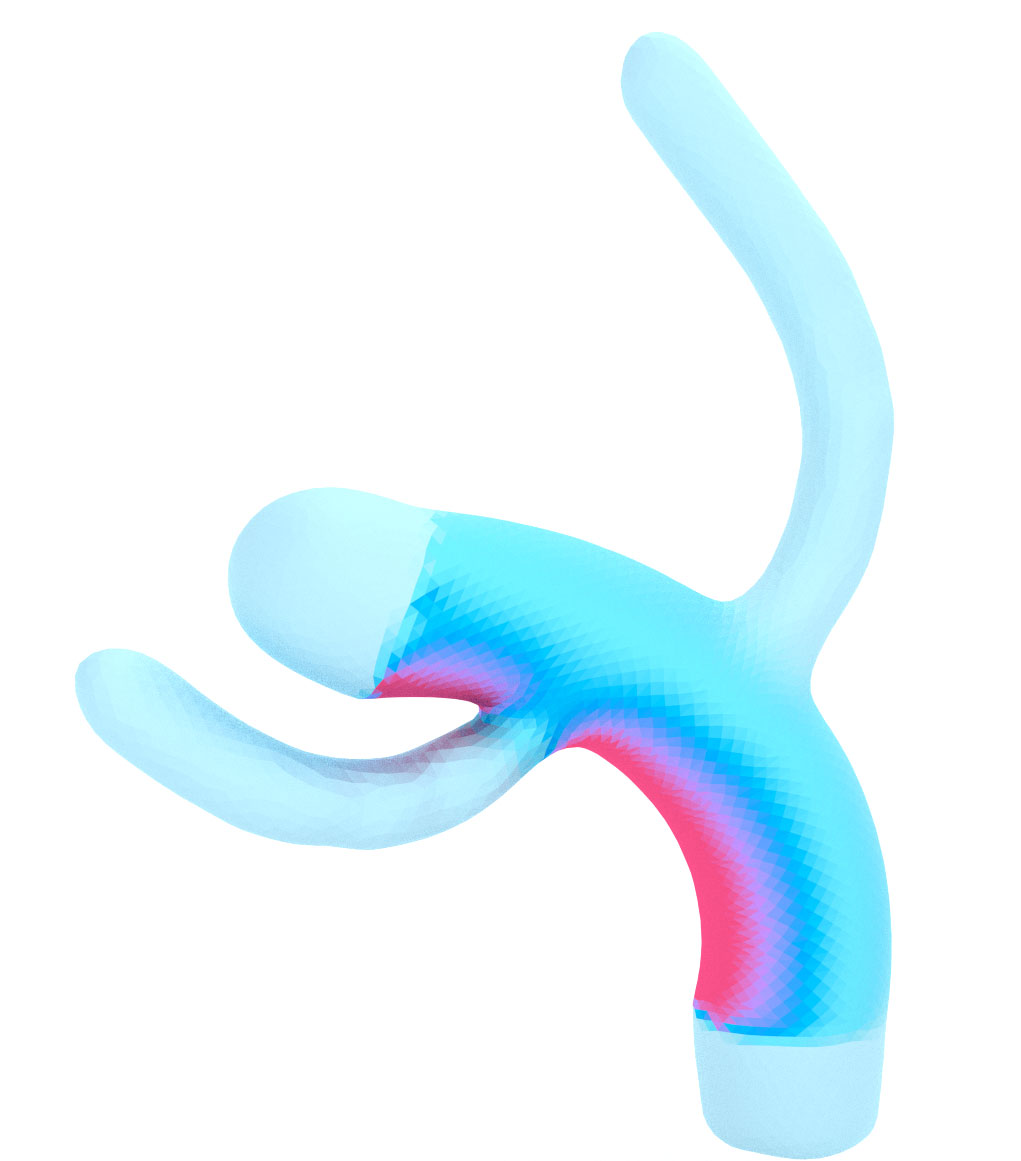} &
\includegraphics[width=0.22\textwidth]{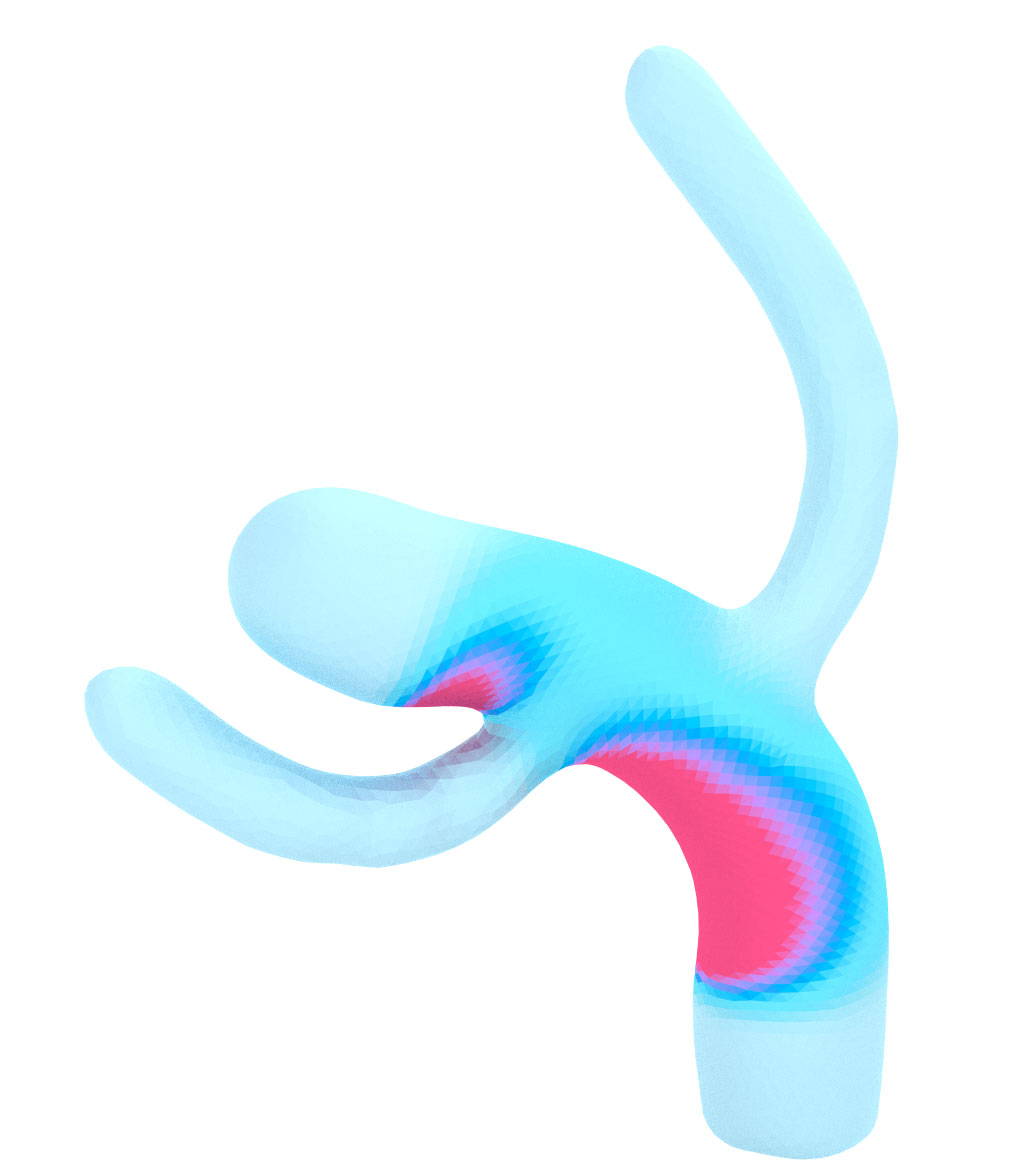} &
\includegraphics[width=0.22\textwidth]{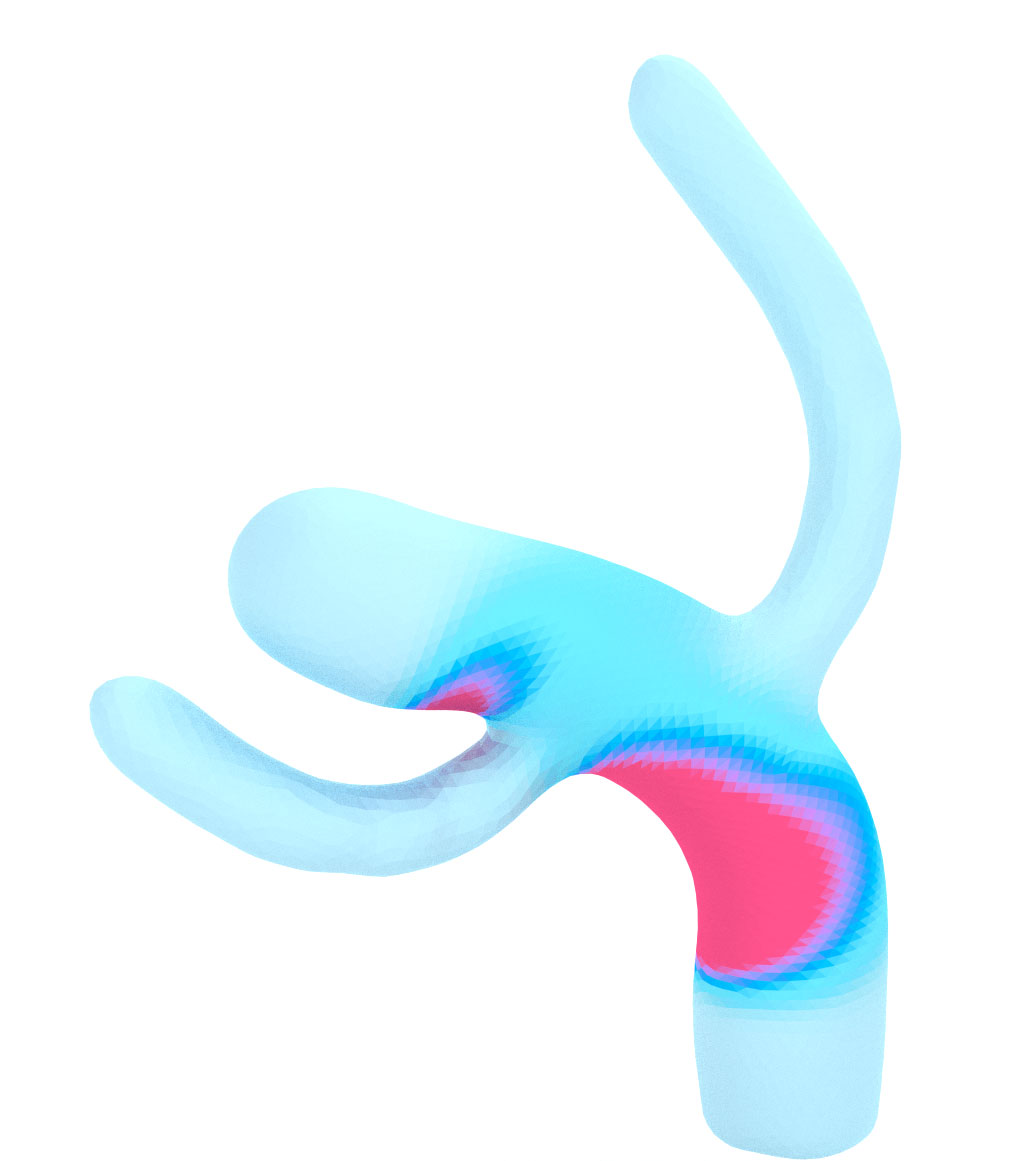}  \\
Zayer \etal \cite{zayer2005harmonic} & $\beta = 0.003$ & $\beta = 0.1$ & $\beta = 0.2$
\vspace{-0.35cm}
	\end{tabular}
	\caption{\emph{Cactus 1}. The entire base and top regions are constrained.
	The base is fixed and the top is translated and rotated. The transition
	between the constrained and free mesh regions is apparent both in the energy
	domain and in the irregularities of mesh normal directions. Low regularization
	weights create smoother transitions and less distortions on the handle
	boundary, while preserving the global shape of the deformation.}
	\label{fig:cactus}
\end{figure*}

\paragraph*{Large Deformation Handles.}
Harmonic surface deformation is commonly used with large deformation handles, as
careful design of these constraints can lead to pleasing deformations.
However, the boundary of large handle regions is also susceptible to artifacts.
The \emph{Cactus 1} model in Figure \ref{fig:cactus} is based on a benchmark
deformation from \cite{botsch2008linear}.
The base of the model is fixed, and the top is rotated and translated.
Using the original method by Zayer \etal \cite{zayer2005harmonic}, the boundary
between the constrained and free mesh regions exhibits strong changes of the
directions of mesh normals, distorting the local shape.
Our regularization reduces distortions of local geometry close to this boundary,
creating a smooth transition between the constrained and free mesh regions.

\paragraph*{Strong Rotations.}
Strong rotations, especially on elongated limbs, are a challenge for a number
of deformation methods \cite{kavan2012articulation}.
Using harmonic surface deformation, strong rotations usually cause loss of
volume.
This effect is illustrated in Figures \ref{fig:cow_face} and \ref{fig:horse}.
The foot of the \emph{Horse} is rotated.
In the resulting deformation, most of the lower leg suffers from volume loss,
creating a ``candy wrapper''-like artifact.
Low values of regularization are effective at preserving some of the limb's
volume to create more plausible results.
In the \emph{Cow} deformation, the horns are rotated upwards while the
front of the face is fixed.
This deformation is even more challenging, as the region deformed is small and
the mesh geometry is rather coarse.
Regularization helps to restore some of the lost volume.
%


\section{Evaluation and Discussion}\label{sec:evaluation}
%
%
%

In this section, we perfom a quantitative evaluation of our approach.
Deformations are created for $\beta \in [0,1)$, using a higher sampling rate for
lower $\beta$ values, for which we usually observe the strongest changes of the
deformation.
Please, see the accompanying \textbf{video} for the deformations at these
$\beta$ values.

\begin{figure}[t]%
\includegraphics[width=\linewidth]{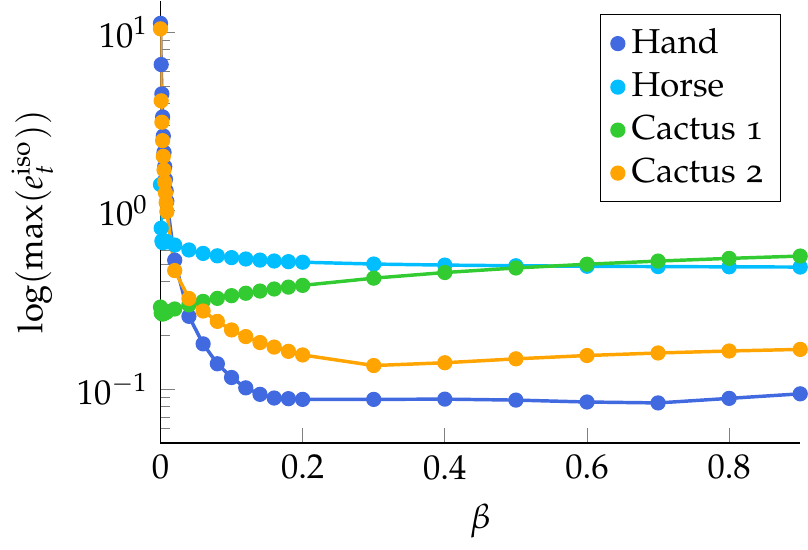}%
\\%
\includegraphics[width=\linewidth]{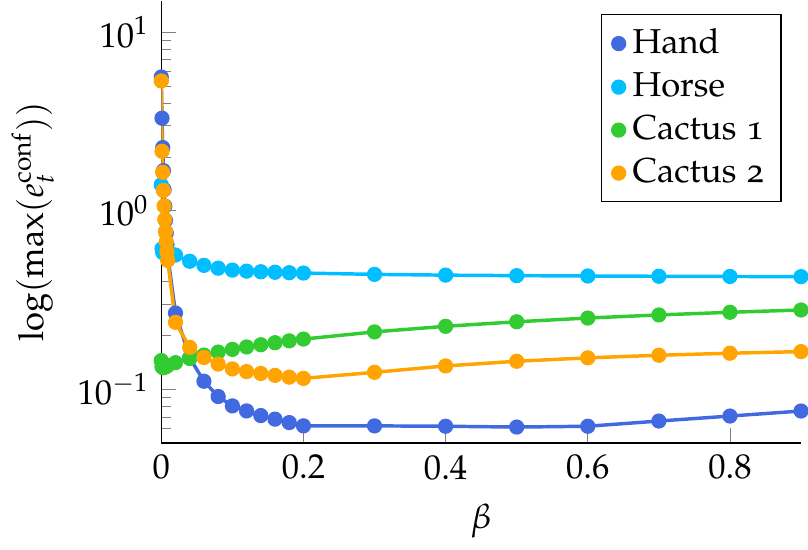}%
\caption{Graphs of the maximal isometric and conformal errors over the
regularization weight $\beta$.  Regularization typically decreases these
errors drastically already for small values of $\beta$. Note the logarithmic
scale on the $y$-axes.}%
\label{fig:max_errors_models}%
\end{figure}

\paragraph*{Maximal Isometric and Conformal Errors.}

Two error measures proved to be most useful for evaluating deformation quality
(see, \eg, \cite{liu2008localglobalparam}):
the local per triangle isometric error $e^{\iso}_t$ is given by the sum of
squared deviations from $1$ of the singular values of the deformation gradient.
The local per triangle conformal error $e^{\conf}_t$ is computed by half of the
squared sum of the pairwise differences between the singular values of the
deformation gradient.
The maximal isometric (conformal) error, $\max(e^{\iso}_t)$
($\max(e^{\conf}_t)$), is given by the maximal value of the isometric
(conformal) error over all triangles.
Both of these error measures indicate strong distortions of the mesh geometry.
They are also loosely related to artifacts such as protruding or intruding
elements, degenerate triangles, and surface self-intersections.
We also examined the total (integrated and normalized) isometric and conformal
errors as possible indicators for deformation quality evaluation, but these
proved ineffective.

Figure \ref{fig:max_errors_models} shows the behavior of the maximal isometric and
conformal errors for different values of $\beta$.
In almost all tested deformations, regularization strongly decreases the maximal
isometric and conformal errors.
%
%
The behavior of these error measurements is different for the \emph{Cactus 1}
example of Figure \ref{fig:cactus}, as they slightly increase with regularization.
The reason is that deformations defined using large handle regions usually
don't result in artifacts associated with large local isometric and conformal
errors.
Additionally, we note that no new local maximal errors are introduced for
moderate $\beta$ values, meaning that deformation quality does not derogate for
higher $\beta$ values.
We also confirm this behavior in all other tested examples, as the total
deformation energy of the regularized deformation defined in
\eqref{eq:deformation_energy_sum} stays within the same order of magnitude as the
original deformation.

\paragraph*{Space of Regularized Deformations.}
We observe that the initial introduction of regularization ($\beta>0$) has
strong effects on deformations suffering from strong artifacts for $\beta=0$.
After this initial reaction interval, regularization creates gradual changes to
the mesh geometry, indicated both by our deformation results (\eg, in
Figures \ref{fig:horse} and \ref{fig:cactus}) and by the behavior of the error measures
$\max(e^{\iso}_t), \max(e^{\conf}_t)$ in Figure \ref{fig:max_errors_models}.
As the regularization energy becomes more dominant with increasing $\beta$
values, $\max(e^{\iso}_t), \max(e^{\conf}_t)$ also slightly increase.
For very high values of $\beta > 0.9$, the regularized energy formulation
strongly deviates from the original problem, which can create new artifacts.
In our experience, choosing a fixed value of $\beta \approx 0.2$ usually
suppresses artifacts effectively without negatively affecting the mesh geometry.
This means that a constant regularization can simply be added to existing
implementations without exposing users to a new parameter.
In addition, although the linear system \eqref{eq:pois_mw_beta} has to be
refactored when $\beta$ changes, examining different $\beta$ values can usually
be done at interactive rates due to the high performance of sparse linear
solvers.
Hence, the space of regularized deformations can also be explored
interactively.

\paragraph*{Curvature-based Differential Operator.}
%
\begin{figure*}
	\centering
	\begin{tabular}{cccc}
\includegraphics[width=0.23\textwidth]{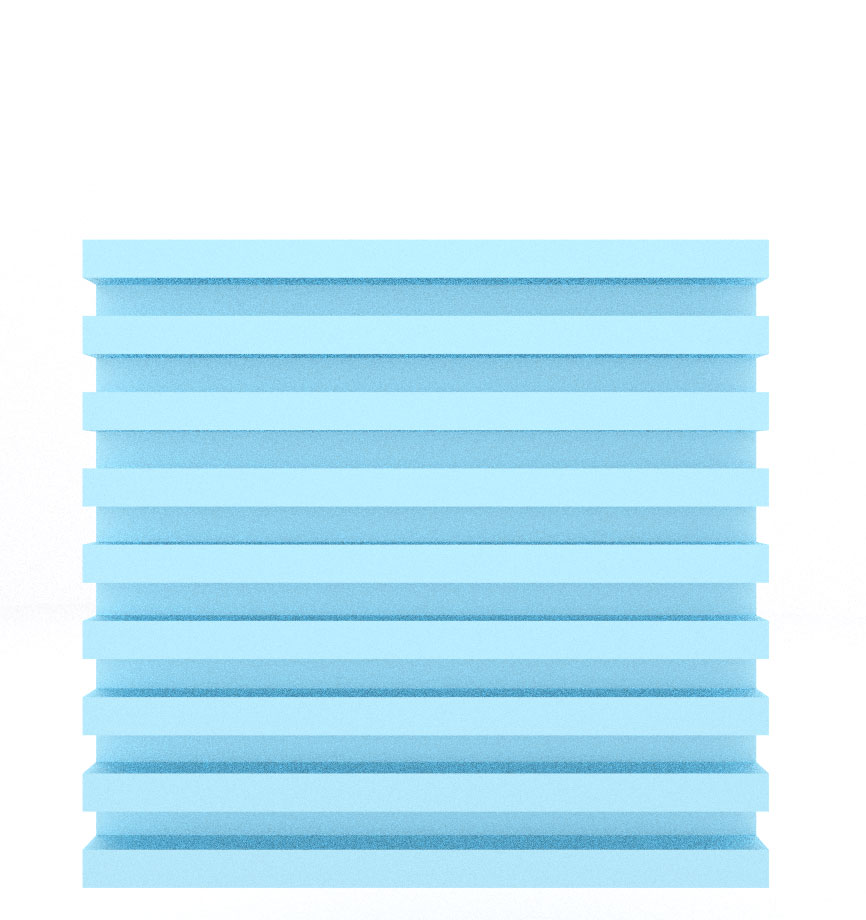} &
\includegraphics[width=0.23\textwidth]{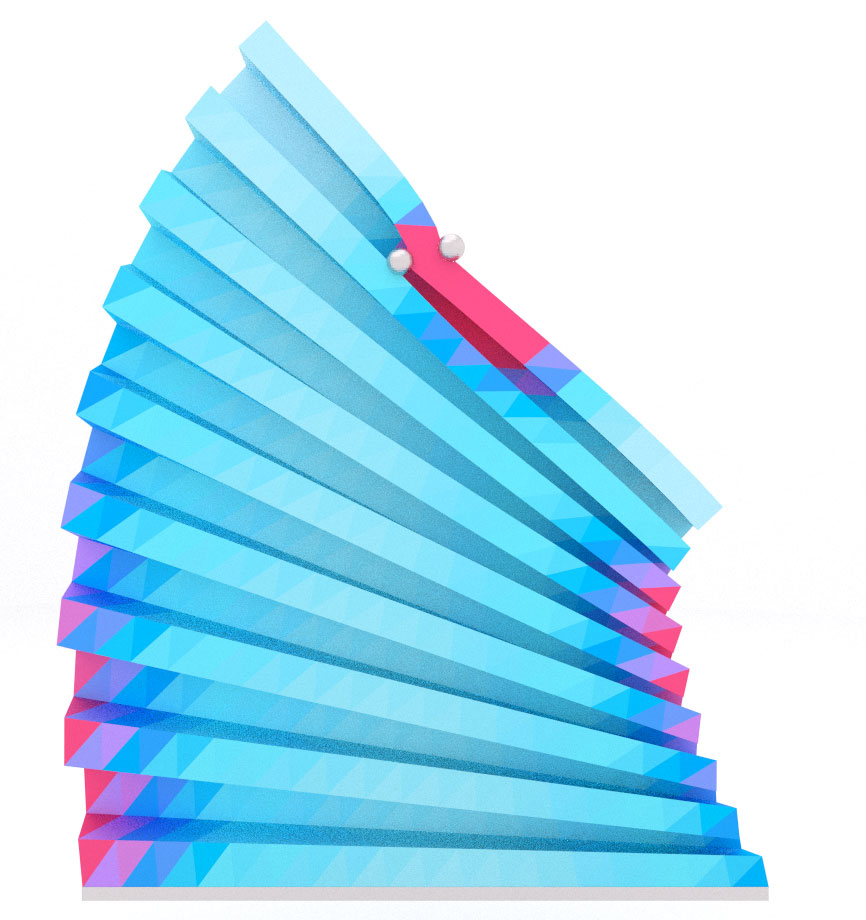} &
\includegraphics[width=0.23\textwidth]{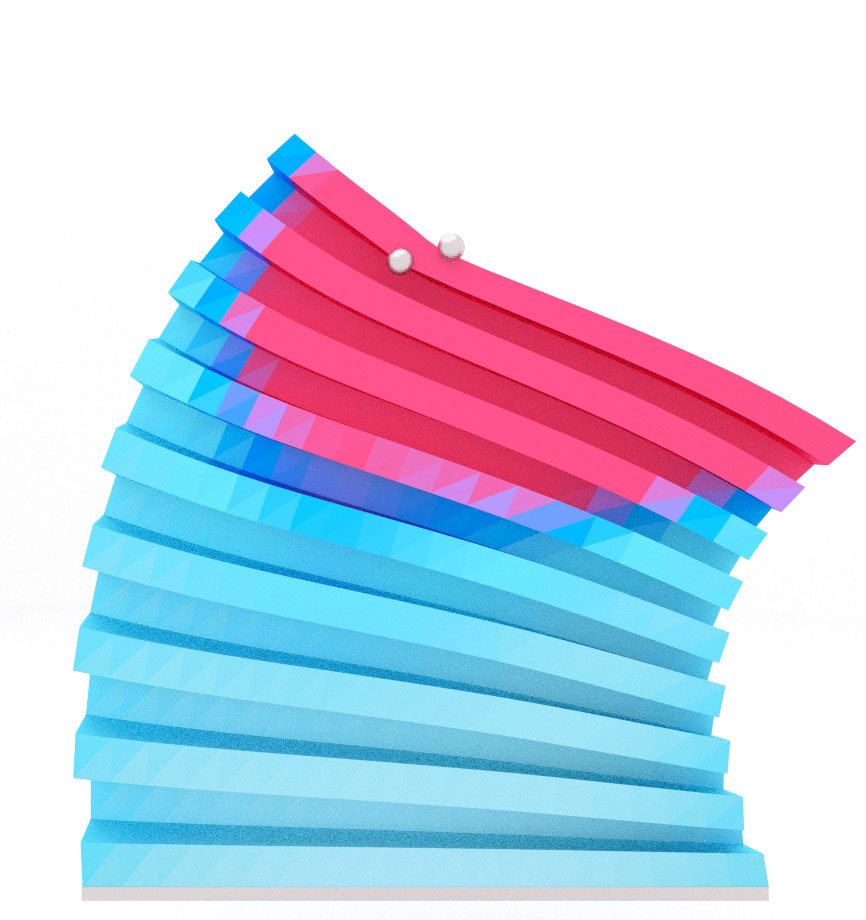} &
\includegraphics[width=0.23\textwidth]{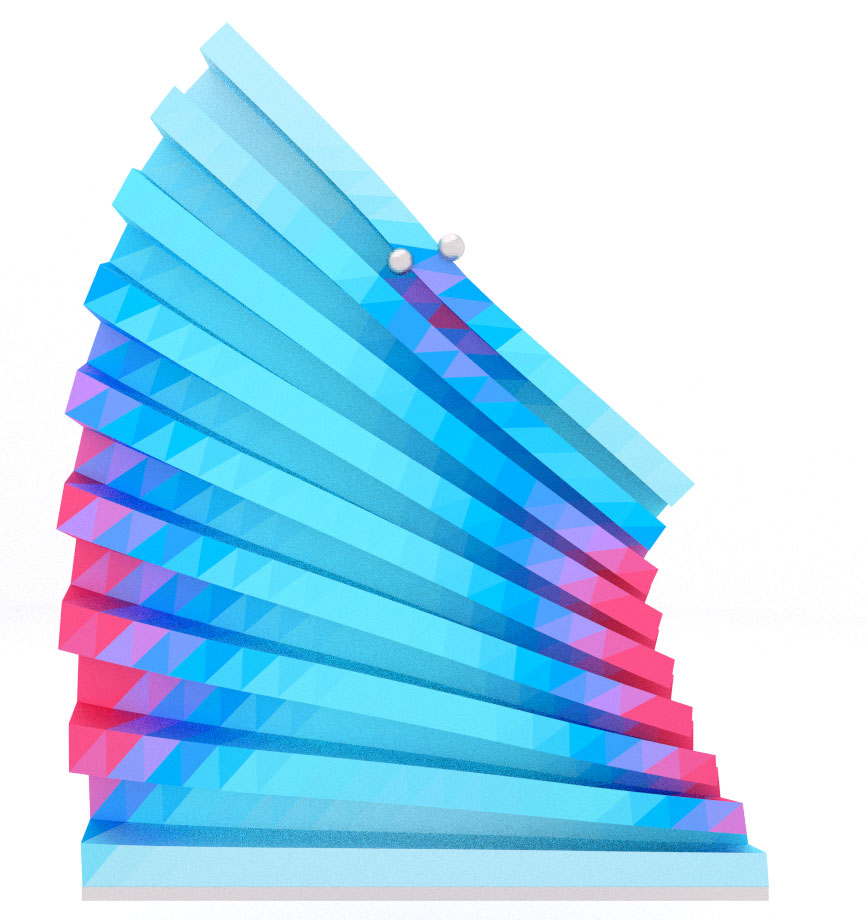}
 \\
Original model & Zayer \etal &  Martinez Esturo   & Ours \\
&  \cite{zayer2005harmonic} &   \etal \cite{janick2014smooth}  & \\
& &  $\beta=0.4$ &  $\beta = 0.4$  \vspace{-0.35cm}
	\end{tabular}
	\caption{\emph{Accordion}. The base is fixed and two vertices on the top row are rotated.
  Harmonic surface regularization \cite{zayer2005harmonic} creates an artifact
  near the deformation handles. Martinez Esturo \etal \cite{janick2014smooth}
  suppress the artifact, but distorts the global mesh shape. Our curvature-based
  differential operator creates a smooth, artifact-free deformation without
  distorting the mesh geometry.}
	\label{fig:accordion}
\end{figure*}

Figure \ref{fig:accordion} demonstrates the benefits of using our curvature-based
energy differential operator on the \emph{Accordion} mesh with highly curved
edges.
The method by Zayer \etal \cite{zayer2005harmonic} suffers from local shape
distortion near the deformation handle.
The differential operator by Martinez Esturo \etal \cite{janick2014smooth}
estimates the energy variation between neighboring triangles ineffectively,
resulting in global shape distortions.
Our curvature-based operator estimates the energy differential more effectively,
resulting in a deformation which has the same geometry as the solution by Zayer
\etal \cite{zayer2005harmonic}, but suppresses the local artifact.
The deviation of Martinez Esturo \etal \cite{janick2014smooth} from our results
increases with higher regularization weight.
For smooth meshes, using our curvature-based differential operator has
negligible effects, as vertex coordinates were only affected marginally.

\paragraph*{Performance.}
Regularization affects the runtime in two ways: it reduces the sparsity of the
linear system \eqref{eq:pois_mw_beta} being solved and it requires a one-time
operation to setup of the curvature-based differential operator $\mD^R$.
Our measurements on an Intel Core i7 $2.2GHz$ system indicates that the effect
of regularization on interactive runtime is insignificant for all tested meshes:
For example, factorization time of the linear system for the \emph{Cactus} model
($\Abs{\cV} = 10k$) is $\approx$ $0.04$ seconds for \emph{both} $\beta = 0$ and
$\beta > 0$, its solution time is $\approx$ $0.003$ seconds.
Similarly, factorization time for the \emph{Hand} model ($\Abs{\cV} = 36.6k$) is
$\approx$ $1.13$ seconds for $\beta = 0$ and $\approx$ $1.14$ seconds for $\beta
> 0$ with a solution time of $\approx$ $0.1$ seconds.
Hence, changing the regularization weight to examine different regularization
weight can be done interactively.

\section{Conclusions}\label{sec:conclusion}


In this work, we have provided an energy regularized formulation of harmonic
surface deformation.
Our approach expands the capabilities of the original method, allowing the
creation of artifact-free deformations for a wider range of deformation
constraints and handle configurations.
Our formulation of a curvature-based differential energy operator improves
the estimation of energy differentials in high-curvature mesh regions.
This reduces geometric distortions introduced by the original energy
differential operator around these regions.
The evaluation of our results demonstrates that even low regularization weights
can effectively suppress many deformation artifacts without negatively affecting
the performance of the original method.
In addition, no new artifacts are created,

\paragraph*{Future work.}
An interesting direction for future work is the application of energy
regularization to other (non-linear) surface deformation approaches,
\eg, to the work of Jacobson \etal \cite{jacobson2012fastskinning}, who observe
similar artifacts.
In addition, since artifacts are usually localized close to handle regions, a
local energy smoothness formulation could achieve better control on deformation
artifacts.

\paragraph*{Acknowledgment.}
The \emph{Horse}, \emph{Hand}, and \emph{Cow} models are provided by the AIM @
Shape project.
The \emph{Cactus} model courtesy of Botsch and Sorkine \cite{botsch2008linear}.

\bibliographystyle{alpha}

\newcommand{\etalchar}[1]{$^{#1}$}


\end{document}